\documentclass[12pt]{article}
\usepackage{epsfig}
\usepackage{graphics}
\usepackage{srcltx}
\usepackage{amsmath}
\usepackage{amsmath,subfigure}
\usepackage{amssymb}
\def\be{\begin{equation}}
\def\ee{\end{equation}}
\def\bp{\begin{pmatrix}}
\def\ep{\end{pmatrix}}
\def\bb{\begin{bmatrix}}
\def\eb{\end{bmatrix}}

\def\bes{\begin{subequations}}
\def\ees{\end{subequations}}

\begin{document}



\begin{center}
{\bf  Families of rational and semi-rational solutions \\ of the partial reverse space-time nonlocal Mel'nikov equation} \\ [11pt]
%{\Large\bf Dynamics of rogue waves \\ on a multi-soliton background \\ in a vector nonlinear Schr\"{o}dinger equation }\\ [11pt]
Wei Liu $^a$, Zhenyun Qin$^b$\footnote{E-mail address: zyqin@fudan.edu.cn}    \\[8pt]
   $^a$ College of Mathematic and Information Science, Shandong Technology and Business University, Yantai,264005, PR China \\ [11pt]
 $^b$ School of Mathematics and Key Lab for Nonlinear Mathematical Models and Methods,  Fudan University, Shanghai 200433, P. R. China\\ [11pt]
%$^c$ Department of Mathematical Sciences, Loughborough University, Loughborough LE11 3TU, United Kingdom
 %\\
\end{center}
\rule{\textwidth}{0.5pt}\\[12pt]

%\author{Wei Liu \and Zhenyun Qin}      %etc.


%\institute{Wei Liu   \at
% College of Mathematic and Information Science, Shandong Technology and Business University, Yantai,264005, PR China  \\
% \email{liuweicc@mail.ustc.edu.cn}
  %\and
%Zhenyun Qin \at
%School of Mathematics and LMNS, Fudan University, Shanghai 200433, PR China \\
%\email{zyqin@fudan.edu.cn} }

%\markboth{Authors' Names}{Instructions for Typing
%Manuscripts (Paper's Title)}

%%%%%%%%%%%%%%%%%%% Publisher's Area please ignore %%%%%%%%%%%%%%%%%%%%%%%
%
%\catchline{}{}{}{}{}
%
%%%%%%%%%%%%%%%%%%%%%%%%%%%%%%%%%%%%%%%%%%%%%%%%%%%%%%%%%%%%%%%%%%%%%%%%%%

%\maketitle

\begin{abstract}
Inspired by the works of Ablowitz, Mussliman and Fokas, a partial reverse space-time nonlocal Mel'nikov equation is introduced. This equation provides two  dimensional analogues of
the nonlocal Schr\"odinger-Boussinesq equation. By employing the Hirota's bilinear method, soliton, breathers and mixed solutions consisting of breathers and periodic line waves are obtained.  Further, taking a long wave limit of these obtained soliton solutions, rational and semi-rational solutions of the nonlocal Mel'nikov equation are derived.
The rational solutions are lumps. The semi-rational solutions are mixed solutions consisting of lumps, breathers and periodic line waves. Under proper parameter constraints, fundamental rogue waves and a semi-rational solution of the nonlocal Schr\"odinger-Boussinesq equation are generated from solutions of the nonlocal Mel'nikov equation.
\end{abstract}

{\normalsize\bf{Key words:}}Partial reverse space-time nonlocal Mel'nikov equation; Nonlocal Schr\"{o}dinger-Boussinesq equation; Hirota's bilinear transformation method; Rational solutions and Semi-rational solutions.
%\PACS{02.30.Jr \and 03.75.Lm \and 04.20.Jb \and 05.45.Yv}%02.30.Jr Partial differential equations 03.65.Ge Solutions of wave equations: %bound states 04.20.Jb	Exact solutions

\section{Introduction}
\label{intro}
It is well known that, nonlinear evolution equations (NLEEs) are well used to describe a variety of nonlinear phenomenon in
fields such as fluids, plasmas, optics, condensed matter physics, particle physics and biophysics \cite{non1,non2,non3,non4,DD1,DD2,DD3,DD4}.
Searching for exact solutions of NLEEs is important in scientific and engineering applications
because it offers a rich knowledge on the mechanism of the complicated physical phenomena modelled by NLEEs.
A variety of powerful methods have been used to obtain solutions to NLLEs and to
investigate the physical properties of these solutions. Examples
of these methods include the Darboux transformation method \cite{dt1,dt2},  the inverse scattering method \cite{in1}, the Hirota bilinear method \cite{hirota}, the homogeneous balance method\cite{hb1,hb2}, the Lie group method \cite{li1,li2}, the
direct method \cite{ma1,ma2,ma3}, and so on \cite{aw1,aw2,aw3}.

 Most of those NLEEs are local equations, namely, the solution's evolution depends only on the local solution value and its local space and time
derivatives. Recently, Ablowitz and Musslimani \cite{ablowitz} introduced the nonlocal nonlinear Schr\"odinger (NLS)  equation
\begin{equation} \label{nNLS}
\begin{aligned}
q_{t}(x,t)-iq_{xx}(x,t)\pm 2iV(x,t)q(x,t)=0\,\,,V(x,t)=q(x,t)q^{*}(-x,t)
\end{aligned}
\end{equation}
which contains the $PT$ symmetric potential $V$.
Fokas extended the nonlocal NLS equation into multidimensional versions, and introduced  the following new integrable nonlocal  Davey-Stewartson (DS) equation  \cite{fokas} :
\begin{equation} \label{DSIeq}
\begin{aligned}
&iA_{t}=A_{xx}+A_{yy}+(\epsilon V-2Q)A,\\
&Q_{xx}-Q_{yy}=(\epsilon V)_{xx}, \\
\end{aligned}
\end{equation}
where
\begin{equation} \label{vv}
\begin{aligned}
&V=A(x,y,t)\,[A(-x,-y,t)]^{*}\,,\epsilon=\pm1,\,
\end{aligned}
\end{equation}
or
\begin{equation} \label{vv1}
\begin{aligned}
&V=A(x,y,t)\,[A(-x,y,t)]^{*}\,,\epsilon=\pm1.
\end{aligned}
\end{equation}
Then a number of new nonlocal integrable equations were proposed and studied \cite{pt1,pt2,pt3,pt4,pt5,pt6,pt7,pt8,pt9,pt10,pt11,jiguang1,jiguang2,pt12,pt13,pt14,pt15}.

However, there are few works of partial reverse space-time nonlocal equations in multidimensional versions.
Inspired by the works of Ablowitz, Mussliman and Fokas, we propose
a partial reverse space-time nonlocal Mel'nikov equation
 \begin{equation} \label{pteq}
\begin{aligned}
&3u_{yy}-u_{xt}-[3u^2+u_{xx}+\kappa\phi\phi^{*}(-x,y,-t)]_{xx}=0,\\
&i\phi_{y}=u\phi+\phi_{xx}.
\end{aligned}
\end{equation}
where $u$ and $\phi$ are functions of $x,y,t$.
Obviously, by replacing $\phi^{*}(-x, y,-t)$ as $\phi^{*}(x,y,t)$ , the nonlocal Mel'nikov equation reduce to the usual Mel'nikov
equation
  \begin{equation} \label{mkeq}
\begin{aligned}
&3u_{yy}-u_{xt}-(3u^2+u_{xx}+\kappa |\phi|^2)_{xx}=0,\\
&i\phi_{y}=u\phi+\phi_{xx}.
\end{aligned}
\end{equation}
Here $u$ is the long wave amplitude (real), $\phi$ is the complex short wave envelope.  As noted by
Mel'nikov \cite{mk1,mk2,mk3,mk4}, this equation may describe an interaction of long waves with short wave packets, and it also could be considered either as a generalization of the KP equation with the addition of a complex scalar field or as a generalization of the NLS equation with a real scalar field \cite{RL}.
Noth that high-order soliton solutions for this equation were derived by Ohta. et.al \cite{jpsj}, rogue wave solutions were derived by Mu and Qin \cite{MQ}, general N-dark soliton solutions of the multi-component Mel'nikov System were investigated by Han. \emph{et. al} \cite{yong}. Besides,
under the variable transformations
 \begin{equation} \label{MStr}
\begin{aligned}
 \xi=x+t,\,\eta=y,\,\tau=t,\,\kappa=-1,
\end{aligned}
\end{equation}
 neglecting the $\tau$-dependence, then rewriting
 $\xi\rightarrow x$ and  $\eta \rightarrow t$,  the nonlocal Mel'nikov equation reduces to
 the nonlocal Schr\"odinger-Boussinesq equation
 \begin{equation} \label{sbeq}
\begin{aligned}
&3u_{tt}-u_{xx}-[3u^2+u_{xx}- \phi\phi^{*}(-x,t)]_{xx}=0,\\
&i\phi_{t}=u\phi+\phi_{xx}.
\end{aligned}
\end{equation}
Rogue wave solutions of the usual Schr\"odinger-Boussinesq equation have been studied by Mu and Qin \cite{MQ2}.

In recent works \cite{jiguang1,jiguang2}, a variety of solutions including $(2+1)$-dimensional breather, rational, semi-rational solutions of the
partially and fully parity-time ($PT$) symmetric nonlocal DS equations have been reported.
Thus it is natural to seek various  exact solutions for the partial reverse space-time nonlocal equations. In this work, we derive families of rational and semi-rational solutions to the partial reverse space-time nonlocal Mel'nikov equation \eqref{mkeq}
by using the Hirota bilinear
method, and then show the key features.

This paper is organized as follows. In section \ref{2}, soliton and breather solutions are derived by employing the Hirota's bilinear method. In section \ref{3}, the main theorem on the rational solutions is provided, and typical features of these rational solutions are shown.  In section \ref{4}, semi-rational solutions consisting of lumps, breathers and periodic line waves are generated, and their unique dynamics are also discussed. Our results are summarized in section \ref{con}.

\section{Soliton, breather solutions of the nonlocal Mel'nikov equation }\label{2}
To using the Hirota bilinear method for constructing soliton solutions
of the partial reverse space-time nonlocal Mel'nikov equation, we consider a transformation different from that considered by
Mu and Qin \cite{mk2}. Here we allow for nonzero asymptotic condition $(\phi,u)\rightarrow(1, 0)$ as $x,y,t\rightarrow\infty$, and look for solution in the form
\begin{equation} \label{uBTtr}
\begin{aligned}
\phi=\frac{g}{f},u=2( {\rm log} f)_{xx},
\end{aligned}
\end{equation}
where  $f\,,g$ are functions with respect to three variables $x\,,y$ and $t$, and satisfy the condition
\begin{equation} \label{ugh}
\begin{aligned}
&f^*(-x,y,-t)=f(x,y,t).
\end{aligned}
\end{equation}
Obviously, $\phi=1, u=0$ is a constant solution of the Equation \eqref{pteq}, and under
the transformation \eqref{uBTtr}, the Equation \eqref{pteq} is cast into the following bilinear form
\begin{equation}\label{ubi}
\begin{aligned}
&(D_{x}^{2}-iD_{y})g \cdot f =0,\\
&(D^{4}_{x}+D_xD_t-3D^2_{y})f \cdot f=\kappa [f^2-gg^{*}(-x,y,-t)].
\end{aligned}
\end{equation}
Here the operator $D$ is the Hirota's bilinear differential operator\cite{hirota} defined by
\begin{equation}
\begin{aligned}
&P(D_{x},D_{y},D_{t}, )F(x,y,t\cdot\cdot\cdot)\cdot G(x,y,t,\cdot\cdot\cdot)\\
=&P(\partial_{x}-\partial_{x^{'}},\partial_{y}-\partial_{y^{'}},
\partial_{t}-\partial_{t^{'}},\cdot\cdot\cdot)F(x,y,t,\cdot\cdot\cdot)G(x^{'},y^{'},t^{'},\cdot\cdot\cdot)|_{x^{'}=x,y^{'}=y,t^{'}=t},\nonumber
\end{aligned}
\end{equation}
where P is a polynomial of $D_{x}$,$D_{y}$,$D_{t},\cdot\cdot\cdot$.

We are in a position to look for soliton solutions of the nonlocal Mel'nikov equation by employing the Hirota's bilinear method, which uses the perturbation expansion \cite{hirota}. To this end, we expand functions $g$ and $f$ in terms of power series of a small parameter $\epsilon$:
\begin{equation}\label{per}
\begin{aligned}
&g=1+\epsilon g_1+ \epsilon^2 g_2 +\epsilon^3 g_3+\cdot\cdot\cdot,\\
&f=1+\epsilon f_1+ \epsilon^2 f_2 +\epsilon^3 f_3+\cdot\cdot\cdot,
\end{aligned}
\end{equation}
Substituting  equation \eqref{per} into bilinear equations \eqref{ubi}, we can obtain soliton solutions $\phi, u$ defined in equation \eqref{uBTtr} with functions
 $f$ and $g$ being the forms of
 \begin{equation}\label{rfg}
\begin{aligned}
f=&\sum_{\mu=0,1}\exp(\sum_{k<j}^{N}\mu_{k}\mu_{j}A_{kj}+\sum_{k=1}^{N}\mu_{k}\eta_{k}),\\
g=&\sum_{\mu=0,1}\exp(\sum_{i<j}^{N}\mu_{k}\mu_{j}A_{kj}+\sum_{k=1}^{N}\mu_{k}(\eta_{k}+i\phi_{k})),\\
\end{aligned}
\end{equation}
where
\begin{equation}\label{cs1}
\begin{aligned}
\eta_{j}&=iP_{j}\,x+Q_{j}\,y+i\Omega_{j}t+\eta_{j}^{0}\,,\exp(i\phi_j)=-\frac{P_j^{4}-Q_{j}^{2}}{P_{j}^{4}+Q_{j}^{2}}+i\frac{2P_{j}^{2}Q_{j}}{P_j^{4}+Q_{j}^{2}},\\
\exp(A_{jk})\\
=&-\frac{k-k\cos(\phi_j-\phi_k)+(P_j-P_k)(\Omega_{j}-\Omega_k)-(P_j-P_k)^4+3(Q_j-Q_k)^2}{k-k\cos(\phi_j+\phi_k)+(P_j+P_k)(\Omega_{j}+\Omega_k)-(P_j+P_k)^4+3(Q_j+Q_k)^2},
\end{aligned}
\end{equation}
and the dispersion relation is
\begin{equation}\label{cs10}
\begin{aligned}
\Omega_{j}P_j(P_j^{4}+Q_{j}^{2})+P_{j}^{4}(P_j^{4}-2Q_{j}^{2}-2k)-3Q_j^4=0\,.
\end{aligned}
\end{equation}
Here $P_{j}\,,Q_{j}\,$ are freely real parameters, and $\eta_{k}^{0}$ is an arbitrary complex constant. The notation $\sum_{\mu=0}$ indicates summation over all possible combinations of $\mu_{1}=0,1\,,\mu_{2}=0,1\,,...,\mu_{N}=0,1$; the $\sum_{j<k}^{N}$ summation is over all possible combinations of the $N$
elements with the specific condition $j<k$.

\begin{figure}[!htbp]
\centering
\subfigure[]{\includegraphics[width=5cm]{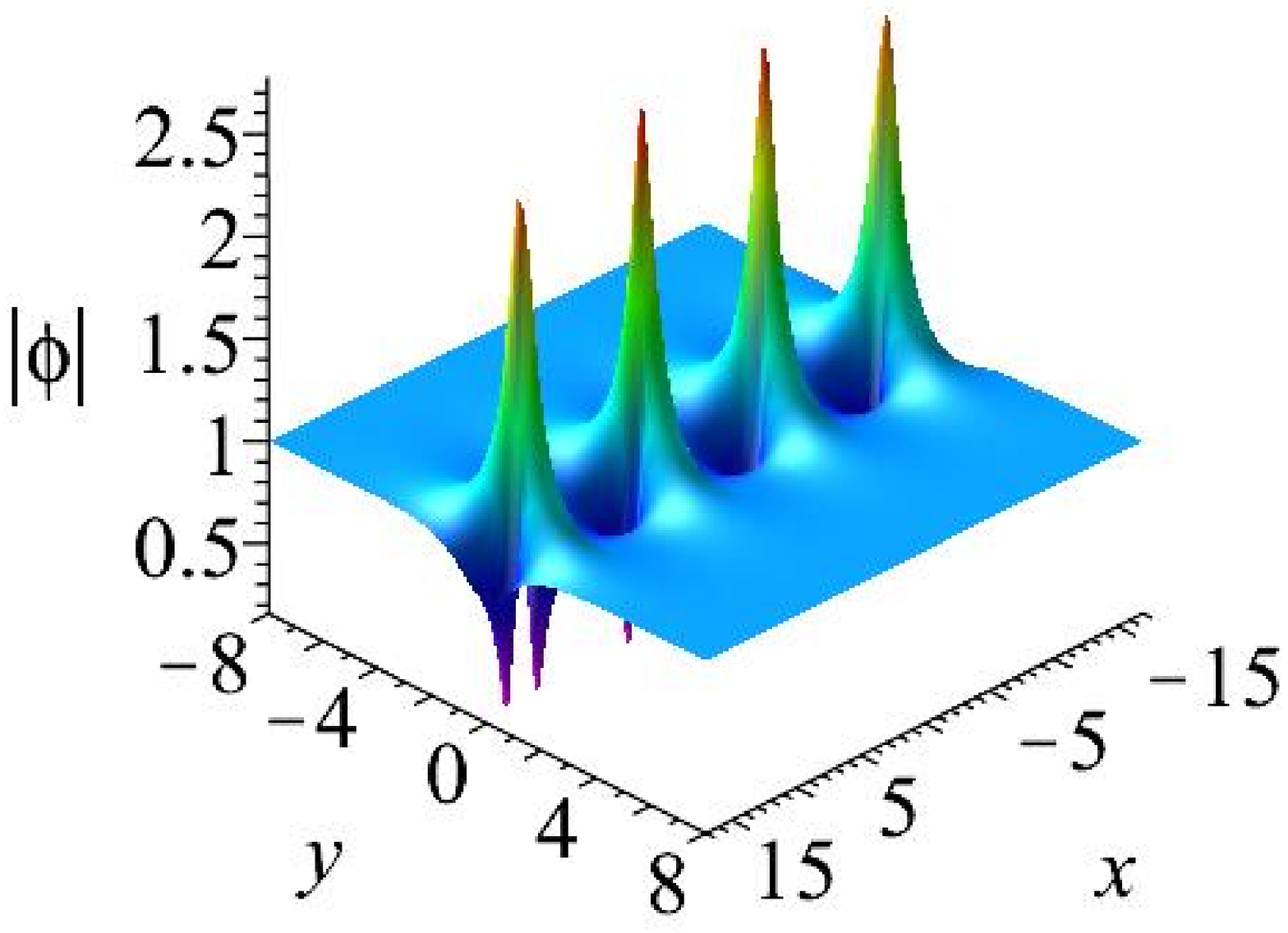}}\quad
\subfigure[]{\includegraphics[width=5cm]{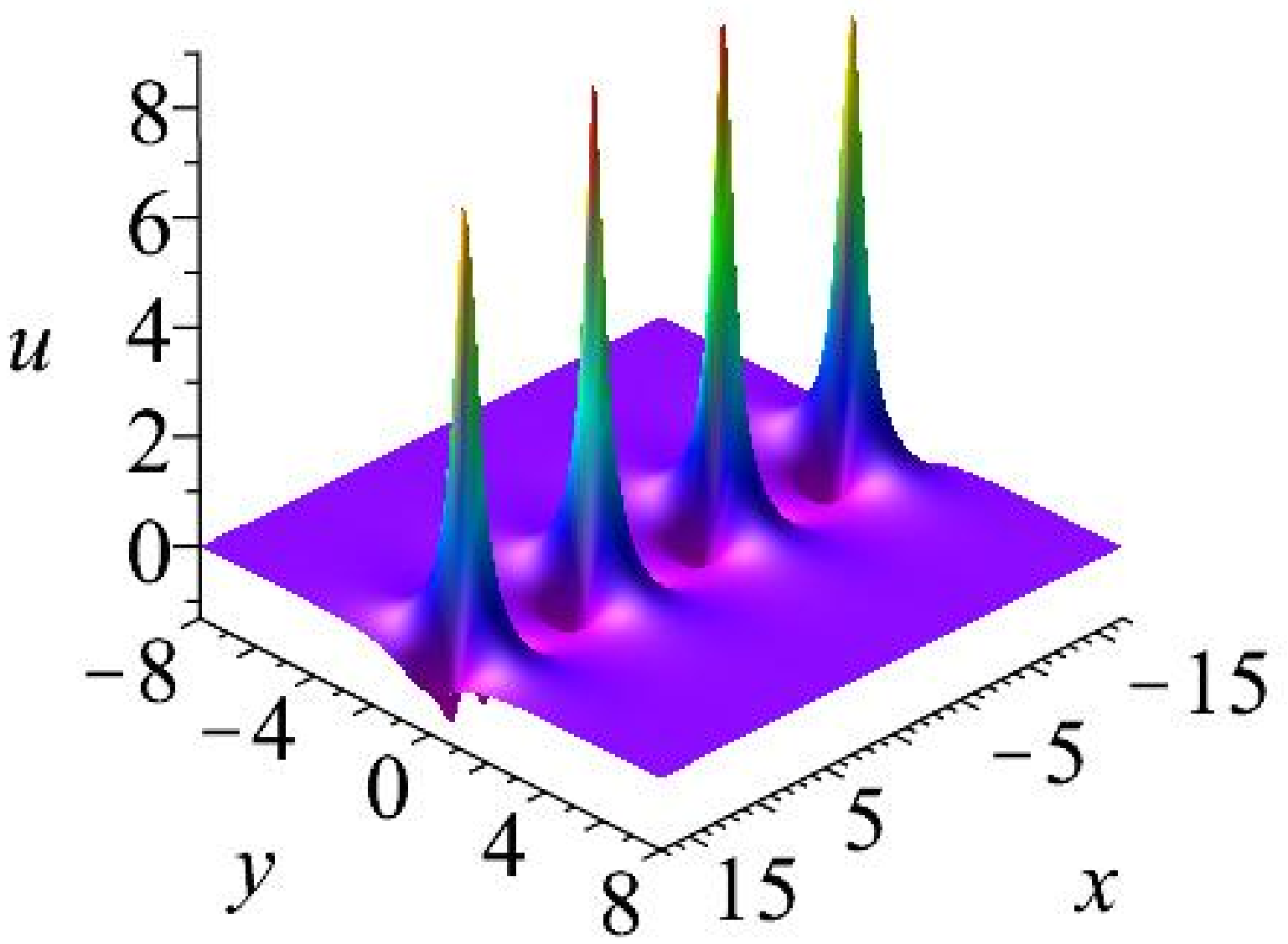}}
\caption{The one-breather solutions $|\phi|$ and $u$ of the nonlocal Mel'nikov equation give by equation \eqref{br} with parameters $\kappa=\frac{1}{2},P=\frac{2}{3}, Q=1, \eta^{0}=0,t=0$.~}\label{fig1}
\end{figure}

\begin{figure}[!htbp]
\centering
\subfigure[]{\includegraphics[width=5cm]{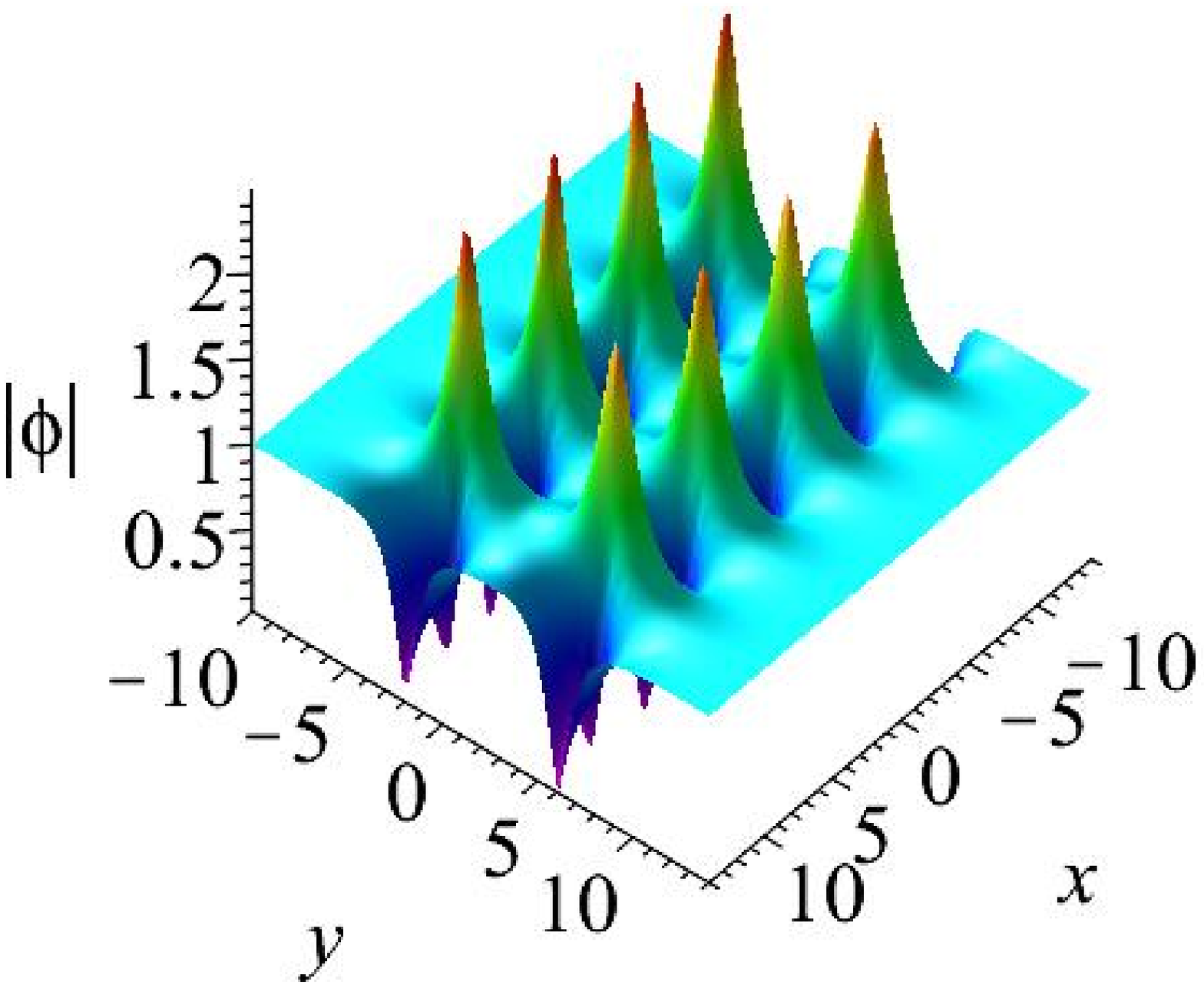}}\quad
\subfigure[]{\includegraphics[width=5cm]{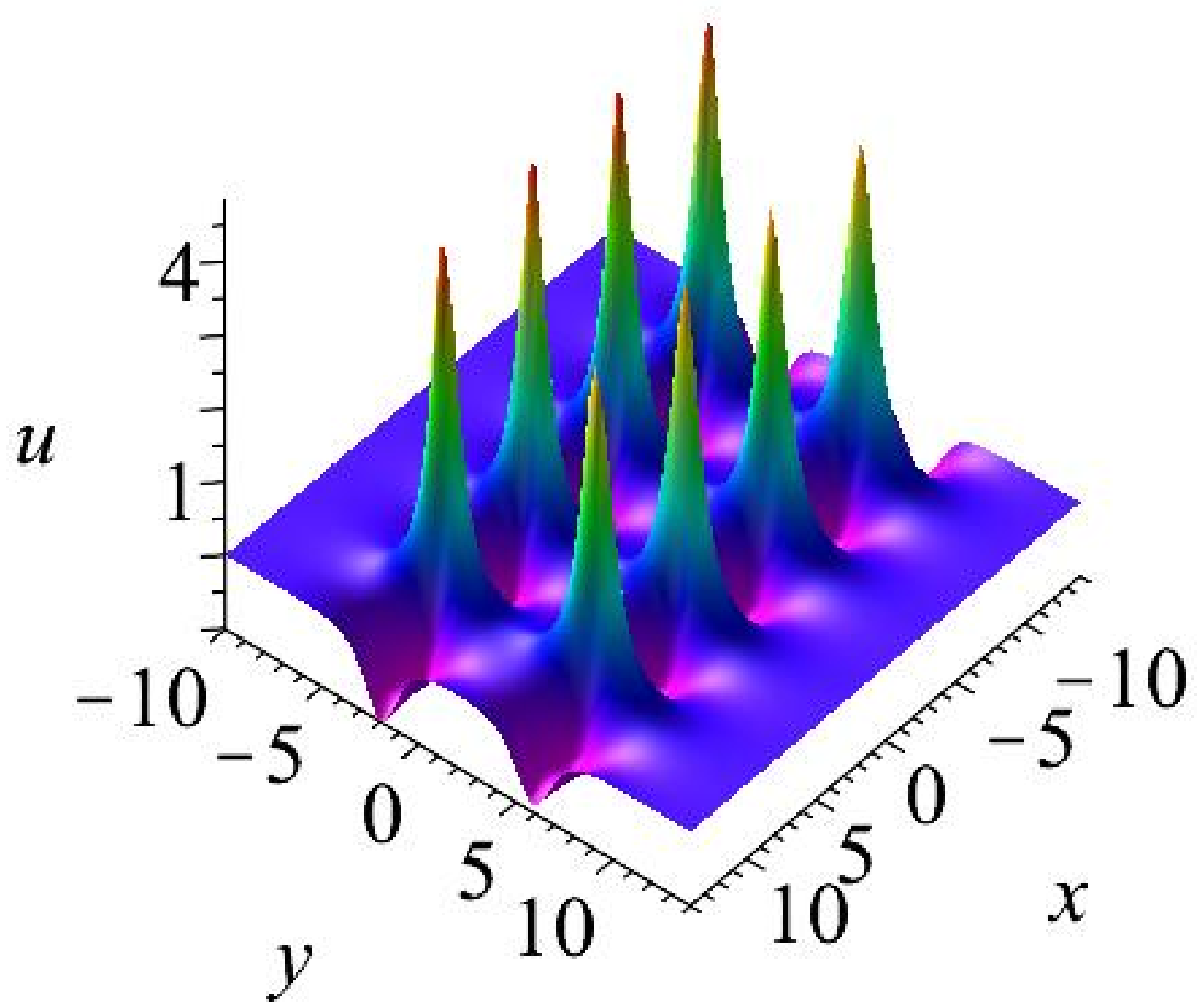}}
\caption{The two-breather solutions $|\phi|$ and $u$ of the nonlocal Mel'nikov equation with parameters $N=4, P_1=1,P_2=-1, P_3=1, P_4=-1, \kappa=\frac{1}{2},Q_2=\frac{2}{3},Q_3=1,Q_4=1, \eta_j^{0}=0\,(j=1,2,3,4),t=0$.~}\label{fig11}
\end{figure}

 In particular, when one takes $Q_{j}=0$ in equation \eqref{cs1}, the corresponding soliton solutions are independent of $y$, and describe periodic line waves periodic in $x$ direction and localized in $y$ direction on the $(x,y)$-plane.
 As soliton solutions of the nonlocal DS equations discussed in Ref.\cite{jiguang1,jiguang2}, these derived solutions of this nonlocal equation also have singularities. However, by suitable constraints of the parameters $P_{j}, Q_{j}, \eta_{j}^{0}$ in equation \eqref{rfg}
 \begin{equation}\label{pa-01}
\begin{aligned}
N=2n\,,P_{n+j}=-P_{j}\,,Q_{n+j}=Q_{j}\,,\eta_{n+j}^{0}=\eta_{j}^{0^{*}},
\end{aligned}
\end{equation}
general nonsingular $n$-breather solutions can be generated.
 For instance, with $N=2$, and parameter choices
  \begin{equation}\label{pc-1}
\begin{aligned}
P_{1}=-P_{2}=P\,,Q_{1}=Q_{2}=Q\,,\eta_{1}^{0}=\eta_{2}^{0}=\eta^{0},
\end{aligned}
\end{equation}
where $P\,,Q\,,\eta^{0}$ are real.
The one breather can also be expressed in terms of hyperbolic and trigonometric functions as
\begin{equation} \label{br}
\begin{aligned}
\phi=\frac{g_{0}}{f_{0}}\,,u=2( {\rm log} f_{0})_{xx},
\end{aligned}
\end{equation}
where
\begin{equation} \label{bfg}
\begin{aligned}
f_{0}=&\sqrt{M}\cosh {\Theta}+\cos(P\,x+\Omega\,t )\,,\\
g_{0}=&\sqrt{M}[\cos^{2}\phi \cosh {\Theta}+\sin^{2}{\phi}\sinh{\Theta}+i\cos\phi\sin\phi(\cosh{\Theta}-\sinh{\Theta})]\\
&+\cos(P\,x+ \Omega t)(\cos\phi+i\,\sin\phi)\,,
\end{aligned}
\end{equation}
$M\,,\Theta\,,\phi$ and $\eta_{0}$ are defined by
 \begin{equation}
\begin{aligned}
&M=1+\frac{P^4}{Q^2} \,,\exp(\eta_{0})=\sqrt{M}\exp(\eta^{0}),\\ &\exp(i\phi)= \frac{Q^2-P^4}{P^4+Q^2}+i \frac{2P^2Q}{P^4+Q^2}\,,\\
&\Theta=-(Q y+\eta_{0}), \Omega=\frac{P^4-2P^4Q^2-2\kappa P^4-3Q^4}{P(P^4+Q^2)}.
\end{aligned}
\end{equation}
The period of $|\phi|$ or $u$ is $\frac{2\pi}{P}$ along the $x$ direction on the $(x,y)$-plane. These two one-breather solutions $|\phi|$ and $u$ are plotted in Fig. \ref{fig1}.  In particular, under parameter constraints
$$
\Omega=P
$$
and the variable transformations defined in equation \eqref{MStr}, then the one-breather solutions defined in equation \eqref{br} reduce to one-breather solutions of the nonlocal Schr\"odinger-Boussinesq equation.  Besides,  the high-order breather solutions can also be generated from equation \eqref{rfg} under parameter constrains \eqref{pa-01}, which still keep periodic in $x$ direction and localized in $y$ direction.  For example, taking parameters
in equation \eqref{rfg}
  \begin{equation}\label{pc-1}
\begin{aligned}
P_{1}=-P_{2},P_3=-P_4, Q_{1}=Q_{2},Q_3=Q_4, \eta_{1}^{0}=\eta_{2}^{*0}, \eta_{3}^0=\eta_{4}^{*0},
\end{aligned}
\end{equation}
the second-order breather solutions $\phi,u$ can be generated, which are illustrated in Fig.\ref{fig11}.

%%%%%%%%%%%%%%%%%%%%%%%%%%%%%%%%%%%%%%%%%%%

\begin{figure}[!htbp]
\centering
\subfigure[]{\includegraphics[width=5cm]{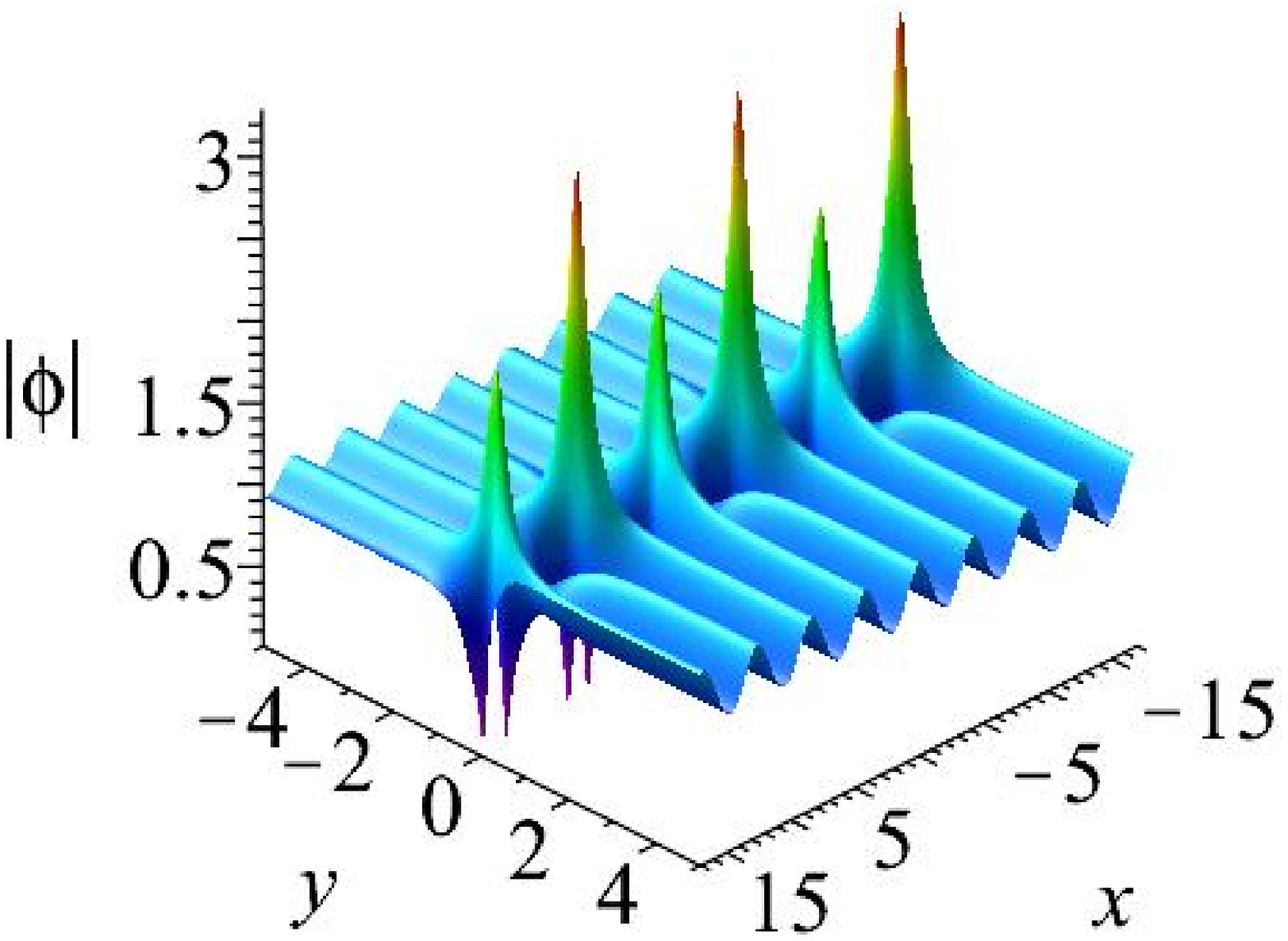}}\quad
\subfigure[]{\includegraphics[width=5cm]{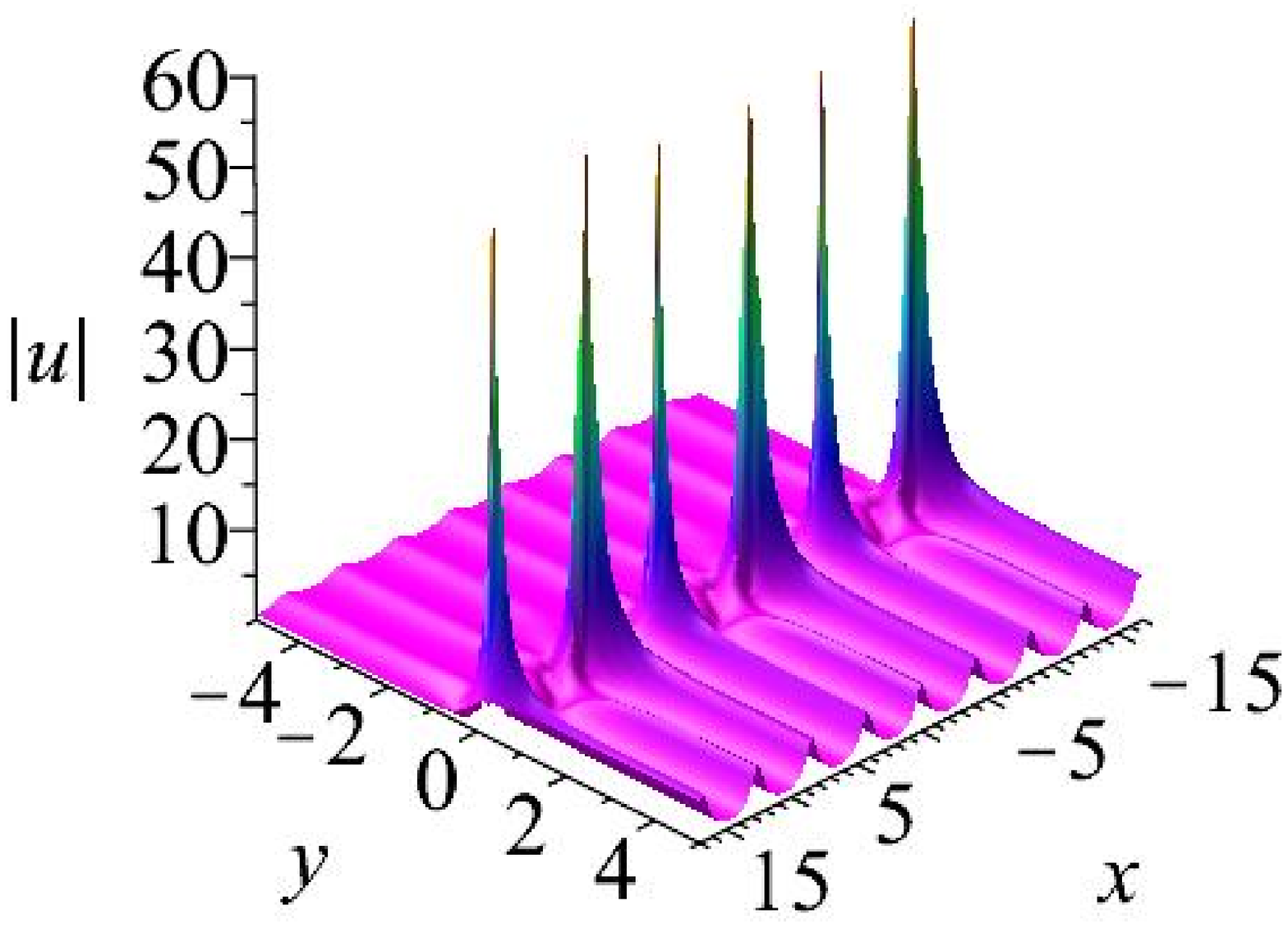}}
\caption{The mixed solutions $|\phi|$ and $|u|$ consisting of one breather and periodic line waves in the nonlocal Mel'nikov equation give by equation \eqref{br} with parameters $\kappa=1,P_0=1, Q_0=2, P3=1, Q_3=0 \eta_1^{0}=0\,,\eta_2^0=0,\,\eta_3^0=-\pi, t=0$.~}\label{fig2}
\end{figure}

In addition to general breather solutions,  mixed solutions consisting of breathers and periodic line waves can also be generated by taking parameters in equation \eqref{rfg}
 \begin{equation}\label{pa-02}
\begin{aligned}
N=2n+1\,,P_{n+j}=-P_{j}\,,Q_{n+j}=Q_{j}\,,Q_{2n+1}=0\,,\eta_{n+j}^{0}=\eta_{j}^{0^{*}}.
\end{aligned}
\end{equation}
This family of hybrid solutions is also nonsingular as breather solutions generated by parameter constraints defined in \eqref{pa-01}, which describes $n$-breather on a background of periodic line waves, and the period of the periodic line waves is $\frac{2\pi}{P_{2n+1}}$. For example,
with $N=3$, and parameter choices in equation \eqref{rfg}
  \begin{equation}\label{pc-1}
\begin{aligned}
P_{1}=-P_{2}=P_0\,,Q_{1}=Q_{2}=Q_0\,,Q_{3}=0, \eta_{1}^{0}=\eta_{2}^{0},
\end{aligned}
\end{equation}
where $P_{0}\,,Q_{0}$ are real, a mixed solution composed of one breather and periodic line waves is obtained, see Fig.\ref{fig2}. The period of the one breather is $\frac{2\pi}{P_0}$, while the periodic line waves is $\frac{2\pi}{P_3}$. In this case, solution $u$ is a complex function, which is different from the solutions of the local Mel'nikov equation, as the later is real.  Again, when one takes parameter constraints
$$
\Omega_1=P_{0},\,\Omega_2=-P_{0}, \Omega_3=P_3
$$
and the variable transformations defined in equation \eqref{MStr}, then corresponding mixed solutions reduce to hybrid solutions consisting of one-breather solutions and periodic line waves to the nonlocal Schr\"odinger-Boussinesq equation.

Besides, another popular method to derive a variety of solutions to soliton equation theoretically is the Darboux  transformation \cite{he1,he2,he3,he4,he5,he6,he7,he8}, but these obtained solutions have demonstrated that the bilinear method is a feasible scheme in computing different types solutions. Examples of these solutions include solitons, breathers, rogue waves, and many other types of rational solutions.  This alternative is especially valuable as most soliton systems possess bilinear forms.

\section{Rational solutions of the nonlocal Mel'nikov equation }\label{3}
To generate rational solutions to the nonlocal Mel'nikov equation, a long wave limit is now taken with the provision
  \begin{equation}\label{long}
\begin{aligned}
\exp(\eta_{j}^{0})=-1,\,( 1\leq j \leq N).
\end{aligned}
\end{equation}
Indeed, under parameter constraints
\begin{equation}
\begin{aligned}
Q_{j}=\lambda_{j}P_{j}\,,\eta_{j}^{0}=i\,\pi\,(1\leq j\leq N),
\end{aligned}
\end{equation}
and taking the limit as $P_{j} \rightarrow 0$ in \eqref{rfg}, then functions $f$ and $g$ in exponential forms are translated into pure rational functions.
 Hence rational solutions to the nonlocal Mel'nikov equation can be presented in the following Theorem.

\noindent\textbf{Theorem 1.} {\sl The partial reverse space-time nonlocal Mel'nikov equation have $Nth-$order rational solutions
\begin{equation} \label{ration}
\begin{aligned}
\phi=\frac{g_{N}}{f_{N}}\,,u = 2( {\rm log} f_{N} )_{xx},
\end{aligned}
\end{equation}
where
\begin{equation}\label{fg}
\begin{aligned}
f_{N}=&\prod_{j=1}^{N}\theta_{j}+\frac{1}{2}\sum_{j,k}^{N}\alpha_{jk}\prod_{l\neq j,k}^{N}\theta_{l}+\cdots\\&+\frac{1}{M!2^{M}}\sum_{j,k,...,m,n}^{N}\overbrace{\alpha_{jk}\alpha_{sl}\cdots\alpha_{mn}}^{M}\prod_{p\neq j,k,...m,n}^{N}\theta_{p}+\cdots,\\
g_{N}=&\prod_{j=1}^{N}(\theta_{j}+b_{j})+\frac{1}{2}\sum_{j,k}^{N}\alpha_{jk}\prod_{l\neq j,k}^{N}(\theta_{l}+b_{l})
+\cdots\\&+\frac{1}{M!2^{M}}\sum_{j,k,...,m,n}^{N}\overbrace{\alpha_{jk}\alpha_{sl}\cdots\alpha_{mn}}^{M}\prod_{p\neq j,k,...m,n}^{N}(\theta_{p}+b_{p})+\cdots,
\end{aligned}
\end{equation}
with
\begin{equation}\label{rt}
\begin{aligned}
\theta_{j}=&i x+\lambda_{j} y-i(3\lambda^2_j+\frac{2k}{\lambda_j^2})t,\\
b_{j}=&\frac{2i}{\lambda_j}, a_{jk}=-\frac{4}{(\lambda_j-\lambda_k)^2}
\end{aligned}
\end{equation}
 the two positive integers $j$ and $k$ are not large than $N$, $\lambda_{j}\,,\lambda_{k}$ are arbitrary real constants, and $\delta_{j}\,,\delta_{k}=\pm1$.
 The first four of \eqref{fg} are written as
 \begin{equation} \label{fgn}
\begin{aligned}
f_{1}=&\theta_{1}\,,\\
f_{2}=&\theta_{1}\,\theta_{2}+a_{12}\,,\\
f_{3}=&\theta_{1}\theta_{2}\theta_{3}+a_{12}\theta_{3}+a_{13}\theta_{2}+a_{23}\theta_{1}\,,\\
f_{4}=&\theta_{1}\theta_{2}\theta_{3}\theta_{4}+a_{12}\theta_{3}\theta_{4}+a_{13}\theta_{2}\theta_{4}+a_{14}\theta_{2}\theta_{3}+a_{23}\theta_{1}\theta_{4}+a_{24}\theta_{1}\theta_{3}\\
&+a_{34}\theta_{1}\theta_{2}+a_{12}a_{34}+a_{13}a_{24}+a_{14}a_{23}\,,\\
g_{1}=&\theta_{1}+b_{1}\,,\\
g_{2}=&(\theta_{1}+b_{1})\,(\theta_{2}+b_{2})+a_{12}\,,\\
g_{3}=&(\theta_{1}+b_{1})(\theta_{2}+b_{2})(\theta_{3}+b_{3})+a_{12}(\theta_{3}+b_{3})+a_{13}(\theta_{2}+b_{2})+a_{23}(\theta_{1}+b_{1})\,,\\
g_{4}=&(\theta_{1}+b_{1})(\theta_{2}+b_{2})(\theta_{3}+b_{3})(\theta_{4}+b_{4})+a_{12}(\theta_{3}+b_{3})(\theta_{4}+b_{4})+a_{13}(\theta_{2}\\
&+b_{2})(\theta_{4}+b_{4})
+a_{14}(\theta_{2}+b_{2})(\theta_{3}+b_{3})+a_{23}(\theta_{1}+b_{1})(\theta_{4}+b_{4})+a_{24}(\theta_{1}\\&
+b_{1})(\theta_{3}+b_{3})+a_{34}(\theta_{1}+b_{1})(\theta_{2}+b_{2})+a_{12}a_{34}+a_{13}a_{24}+a_{14}a_{23}\,.\nonumber
\end{aligned}
\end{equation}
The above formulae for $f$ and $g$ will be used to express explicit form of rational solutions.
}

\noindent\textbf{Remark 1.} These rational solutions can be classified into two patterns:

(1) By restricting the parameters
\begin{equation}\label{constrain}
\begin{aligned}
N=2n, \lambda_{n+j}=-\lambda_{j}\,(j=1,2,3,\cdot\cdot\cdot n)
\end{aligned}
\end{equation}
in Theorem 1, the corresponding rational solutions are nonsingular, which are $n$th-order lumps.

(2) When the parameters satisfy parameter constraints defined in \eqref{constrain}
the corresponding  have singularity at point $(x,y,t)=(0,0,0)$.  Thus hereafter we just focus on nonsingular rational solutions under parameter constraints
defined in \eqref{constrain}.

 To demonstrate the typical dynamics of these nonsingular rational solutions,  we first consider the first-order lump solutions. Indeed, taking
\begin{equation}\label{xx1}
\begin{aligned}
N=2, \lambda_{2}=-\lambda_1=\lambda
\end{aligned}
\end{equation}
in formula \eqref{fg}, the one-lump solutions can be generated. In this case, solutions $u$ and $\phi$ can be expressed
\begin{equation}\label{1-ra}
\begin{aligned}
\phi&=1+\frac{g_2^0}{f_2}\\
&=1+\frac{4iy\lambda^2-4}{(x\lambda+(2k\lambda^{-1}+3\lambda^3)t)^2+\lambda^4y^2+1},\\
u &= 2( {\rm log} f_{2} )_{xx}\\
&=4\frac{(x\lambda^2-(2k+\lambda^4)t)^2-\lambda^6y+\lambda^2}{[((x\lambda+(2k\lambda^{-1}+3\lambda^3)t)^2+\lambda^4y^2+1)^2+\lambda^4y^2+1]^2}.
\end{aligned}
\end{equation}
As discussed in Ref. \cite{jiguang2}, solutions $\phi, u$ are constant along the $[x(t), y(t)]$ trajectory where
$$
x-\frac{2k+\lambda^4}{2\lambda^2}t=0,\,y=0.
$$
Besides, at any fixed time, $(\phi,u)\rightarrow (1,0)$ when $(x, y)$ goes to infinity. Hence these rational solutions are permanent lumps moving on the constant backgrounds.

\begin{figure}[!htbp]
\centering
\subfigure{\includegraphics[width=5cm]{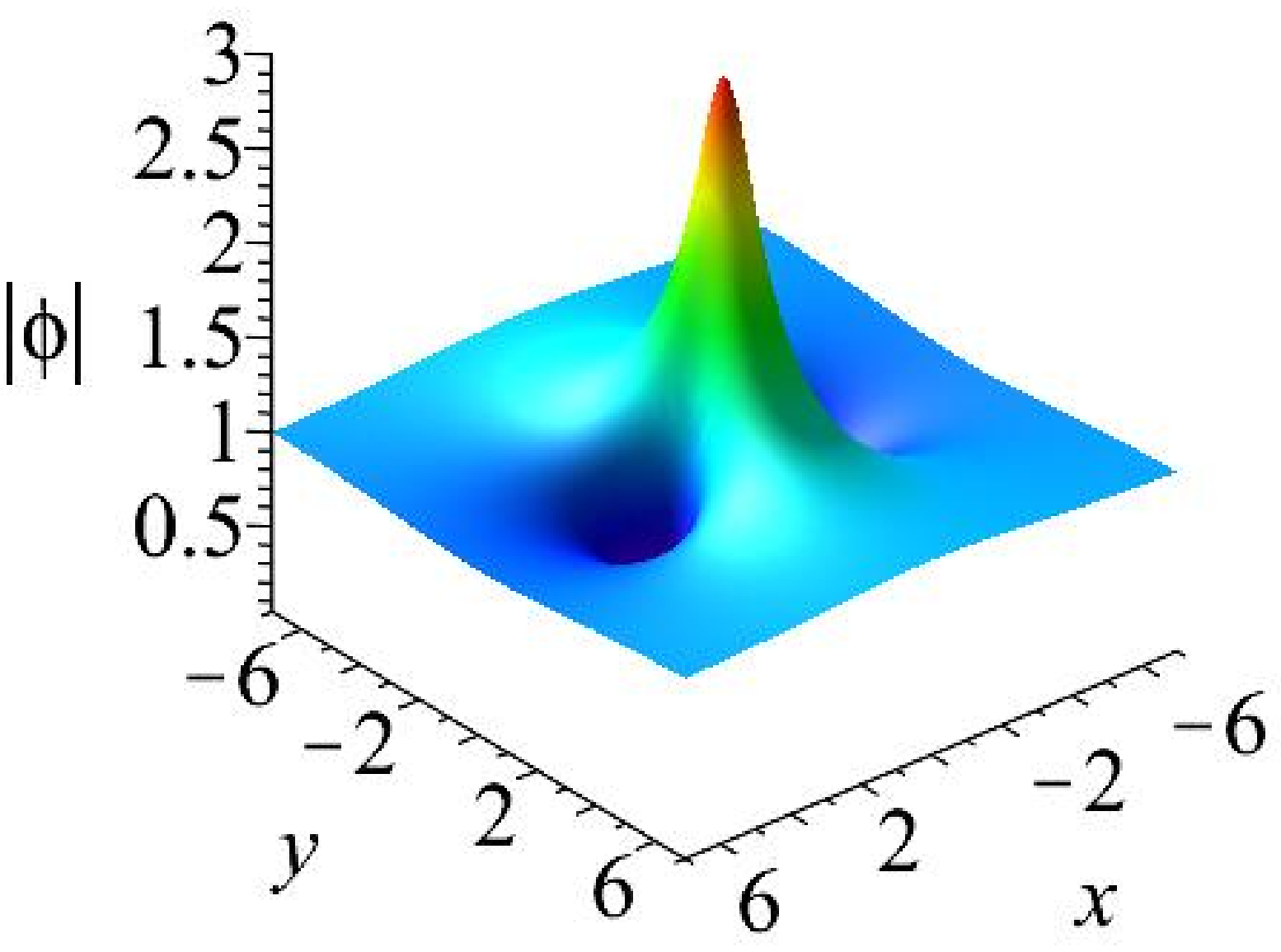}}\quad
\subfigure{\includegraphics[width=5cm]{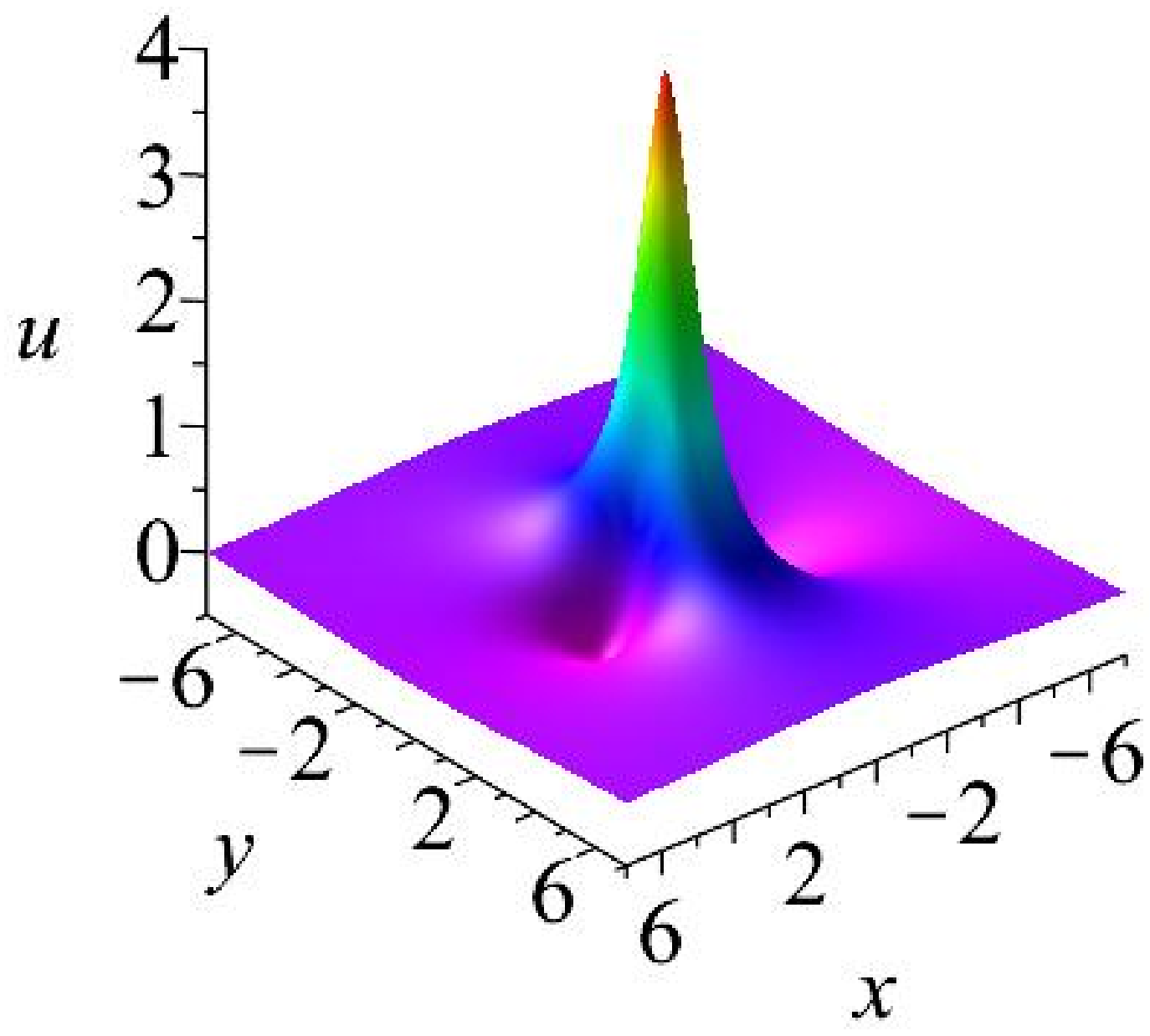}}
\caption{The one-lump solutions $|\phi|$ and $|u|$ of the nonlocal Mel'nikov equation given by equation \eqref{1-ra} with parameters $\kappa=1,\lambda=1,t=0$.~}\label{fig3}
\end{figure}

After a shift of time, the patterns of lump  solutions  do not change. Thus, we can discuss of the patterns of the lump solutions at $t=0$ without loss of generality.
In this case, the two solutions $\phi, u$ have critical points
$$
A_1(x,y)=(0,0), A_2(x,y)=(\frac{\sqrt{3}}{\lambda},0), A_3(x,y)=(-\frac{\sqrt{3}}{\lambda},0),
$$
which are derived from the first order derive $\frac{\partial |\phi| }{\partial x}= \frac{\partial |\phi| }{\partial y}=0 \, (\frac{\partial u}{\partial x}= \frac{\partial u }{\partial y}=0)$. The second derivatives at the these three critical points are given by
\begin{equation}
\begin{aligned}
&|\psi|_{xx} |_{A_1}=-48\lambda^2,  \\
&(|\psi|_{xx}|\psi|_{yy}-|\psi|^2_{xy} )|_{A_2}=(|\psi|_{xx}|\psi|_{yy}-|\psi|^2_{xy} )|_{A_3}=768\lambda^6,\\
&u_{xx} |_{A_1}=-24\lambda^4,\\
&(u_{xx}u_{yy}-u^2_{xy}) |_{A_2}= (u_{xx}u_{yy}-u^2_{xy})  |_{A_3}=\frac{41}{2}\lambda^{10}.
\end{aligned}
\end{equation}
Hence, point $A_1$ is the maximum value point of solutions $\phi$ and $u$, and points $A_2,A_3$ are the minimum value points. That fact indicates that there are only bright lumps in the nonlocal Mel'nikov equation, which is different from the local Mel'nikov equation \cite{mk2}, as the later possesses three patterns of lumps. Besides, by comparison of this rational solutions of the local Mel'nikov equation, it is obvious that the expression of $g_2^{0}$ in \eqref{1-ra} only contains the variable $y$, but the corresponding solutions contain all the variables $x,y$ and $t$ in the local Mel'nikov equation. Thus their solutions are different in the expressions.
 The one-lumps solutions $u$ and $\phi$ are plotted in Fig. \ref{fig3}.  In particular, when one takes
\begin{equation}\label{xx2}
\begin{aligned}
\lambda=\frac{\sqrt{6}}{3},\kappa=-1,
\end{aligned}
\end{equation}
in equation \eqref{1-ra}, and then taking the variable transformations defined in equation \eqref{MStr}, the two dimensional lumps solutions \eqref{1-ra} reduce to rogue wave solutions of the nonlocal Schr\"odinger-Boussinesq equation, which can be expressed as
\begin{equation}\label{rw-1}
\begin{aligned}
\psi&=1-\frac{24it+36}{6x^2+4t^2+9},\\
u&=\frac{24(4t^2+9-6x^2)}{(6x^2+4t^2+9)^{2}}.
\end{aligned}
\end{equation}
$|\psi|$ reaches maximum amplitude 3 (i.e., three times the background
amplitude) at the point $(0,0)$, and the minimum amplitude $0$ at points $(\frac{3\sqrt{2}}{2},0)$ and $(-\frac{3\sqrt{2}}{2},0)$.
$u$ reaches maximum amplitude $\frac{8}{3}$ at the point $(0,0)$, and the minimum amplitude $-\frac{1}{3}$ at points $(\frac{3\sqrt{2}}{2},0)$ and $(-\frac{3\sqrt{2}}{2},0)$. Note that the height
of the background of $u$ is $0$.

Higher-order lumps can be derived with larger $N = 2n (n \geq 2)$ and other parameters meet the parameter constraints
defined in \eqref{constrain}, which describe the interaction of $n$ individual fundamental lumps. For example,
taking
\begin{equation} \label{pa-3}
\begin{aligned}
N=4\,,\lambda_{1}=-\lambda_{3},\lambda_{2}=-\lambda_{4},\delta_{1}\delta_{3}=-1,
\end{aligned}
\end{equation}
 the second-order lump solutions can be obtained from \eqref{ration}, which is
\begin{equation} \label{2-ration}
\begin{aligned}
\phi=\sqrt{2}\frac{g_{4}}{f_{4}}\,,u=2( {\rm log} f_{4} )_{xx}.
\end{aligned}
\end{equation}
The explicit form of solution  $\phi$ and $u$ with parameters
\begin{equation} \label{xx4}
\begin{aligned}
\lambda_1=-\lambda_2=2,\lambda_3=-\lambda_4=1,k=1
\end{aligned}
\end{equation}
can be expressed as
\begin{equation} \label{2-ra}
\begin{aligned}
\phi=1+\frac{g_4^0}{f_4},\,u=2( {\rm log} f_{4} )_{xx},\,
\end{aligned}
\end{equation}
where
\begin{equation} \label{xx5}
\begin{aligned}
f_4=&{x}^{4}(36)+{x}^{3}(-1260t)+ {x}^{2}(15525t^{2}+36y^{2}+365)\\
&+x(-78750t^{3}-6590t-2340ty^{2})\\
&+(140625t^{4}+25850t^{2}+9225t^{2}y^{2}+144y^{4}-359y^{2}+\frac{5329}{9}),\\
g^{0}_{4}=&36[(110\,tx-650\,{t}^{2}-5\,{x}^{2}-33\,{y}^{2}+{\frac {146}{9}})+iy(725\,{t}^{2}+8\,{x}^{2}\\
&-140\,tx+20\,{y}^{2}-{\frac
{455\,}{9}})].
\end{aligned}
\end{equation}
\begin{figure}[!htbp]
\centering
\subfigure{\includegraphics[width=5cm]{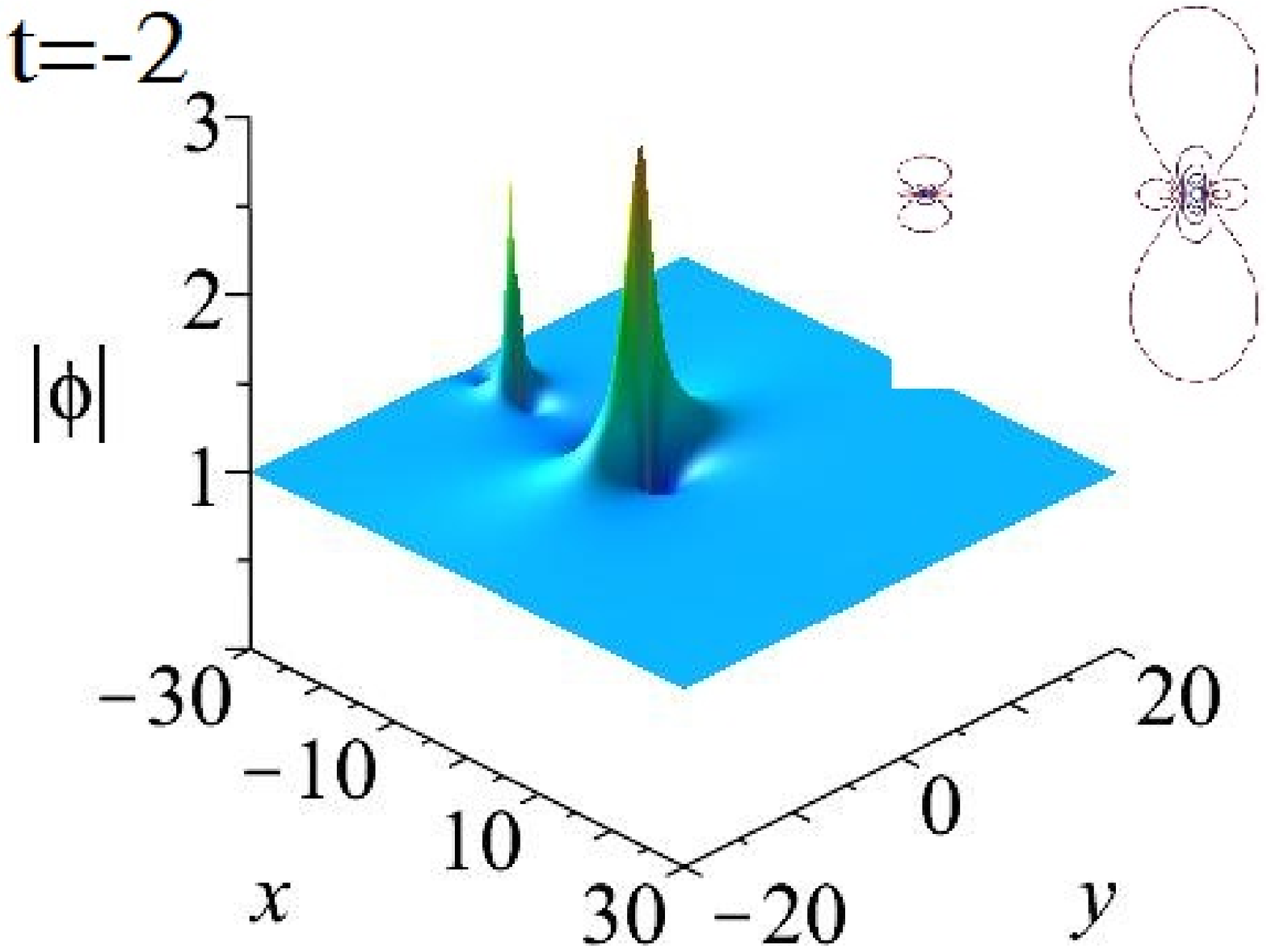}}\quad
\subfigure{\includegraphics[width=5cm]{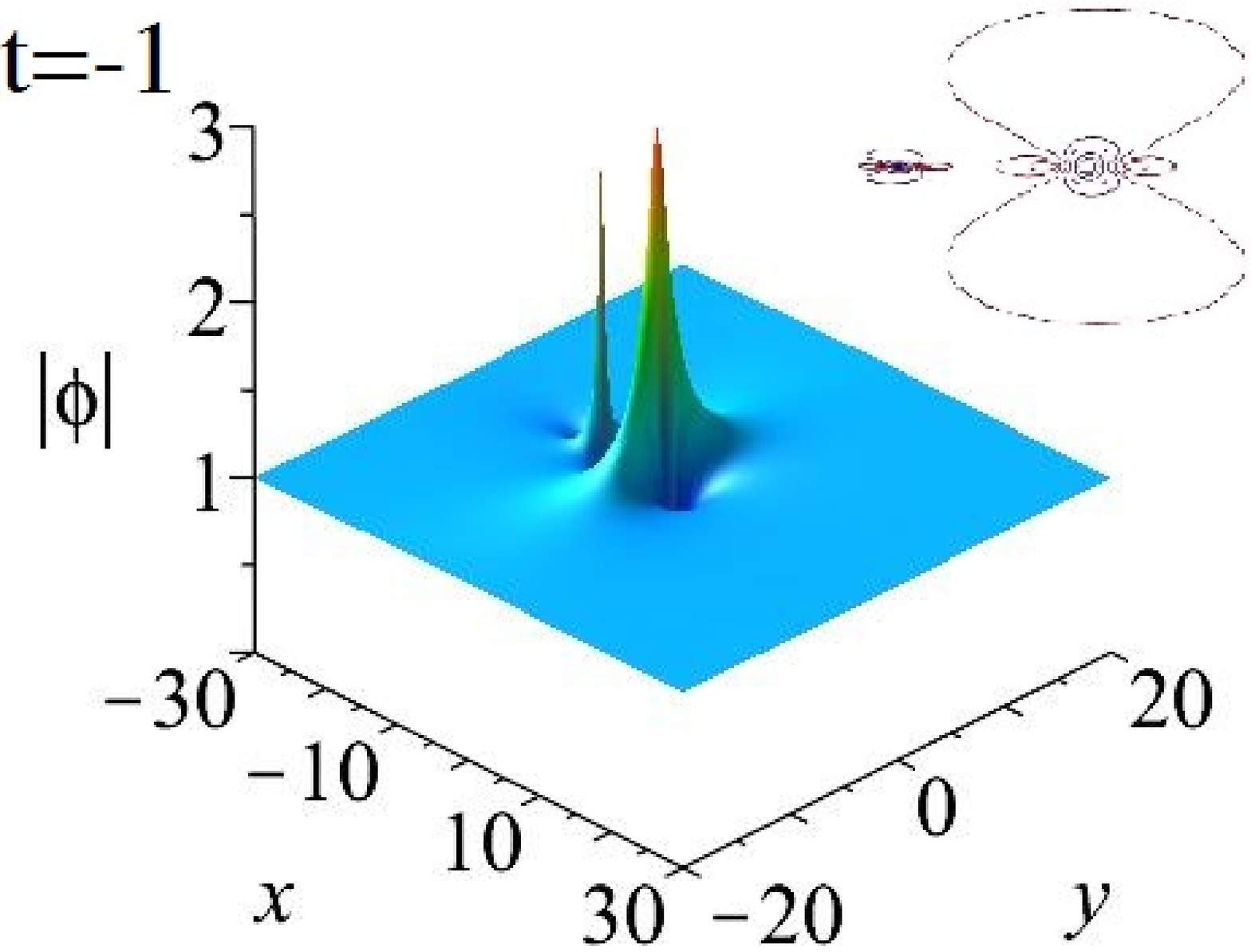}}
\subfigure{\includegraphics[width=5cm]{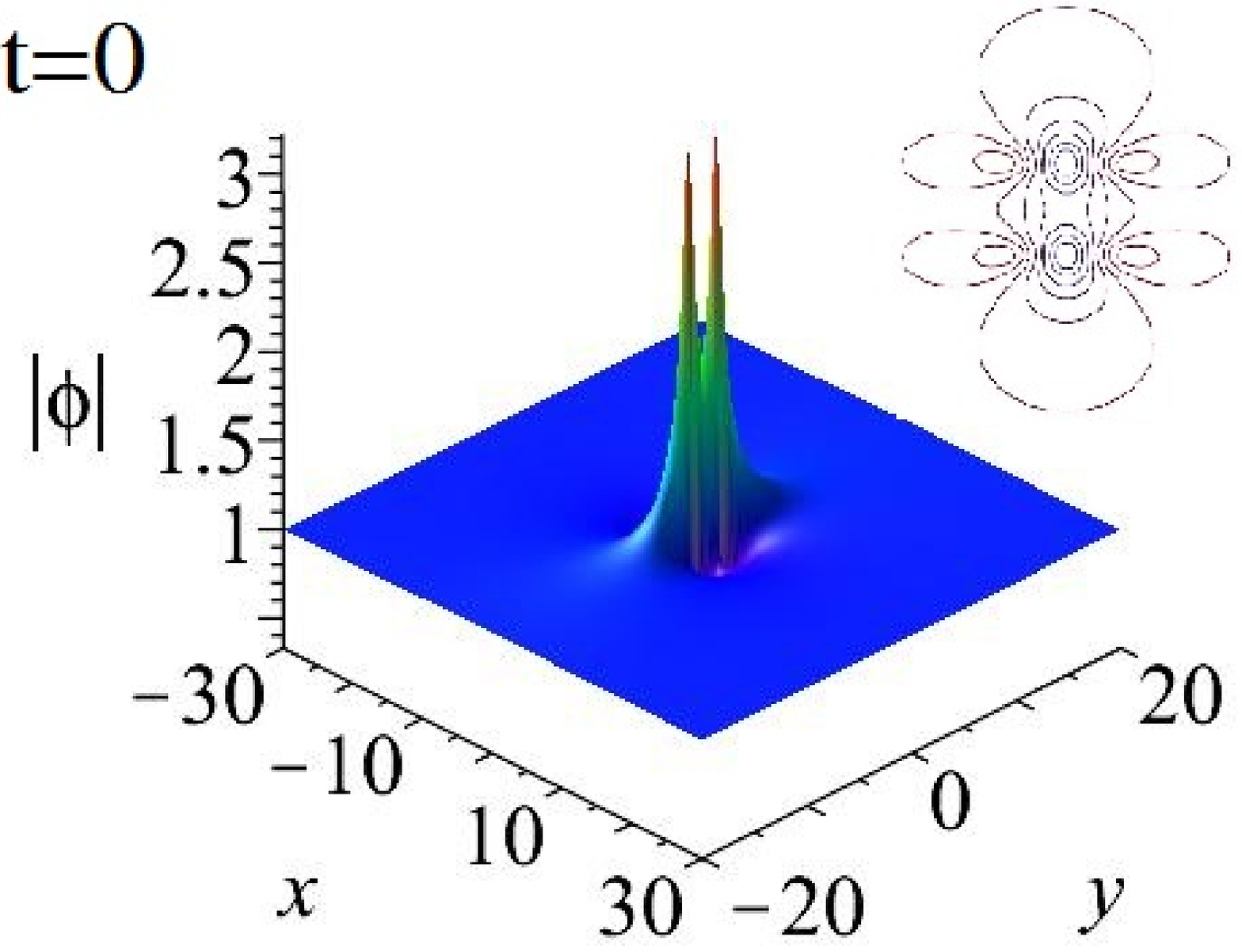}}\quad
\subfigure{\includegraphics[width=5cm]{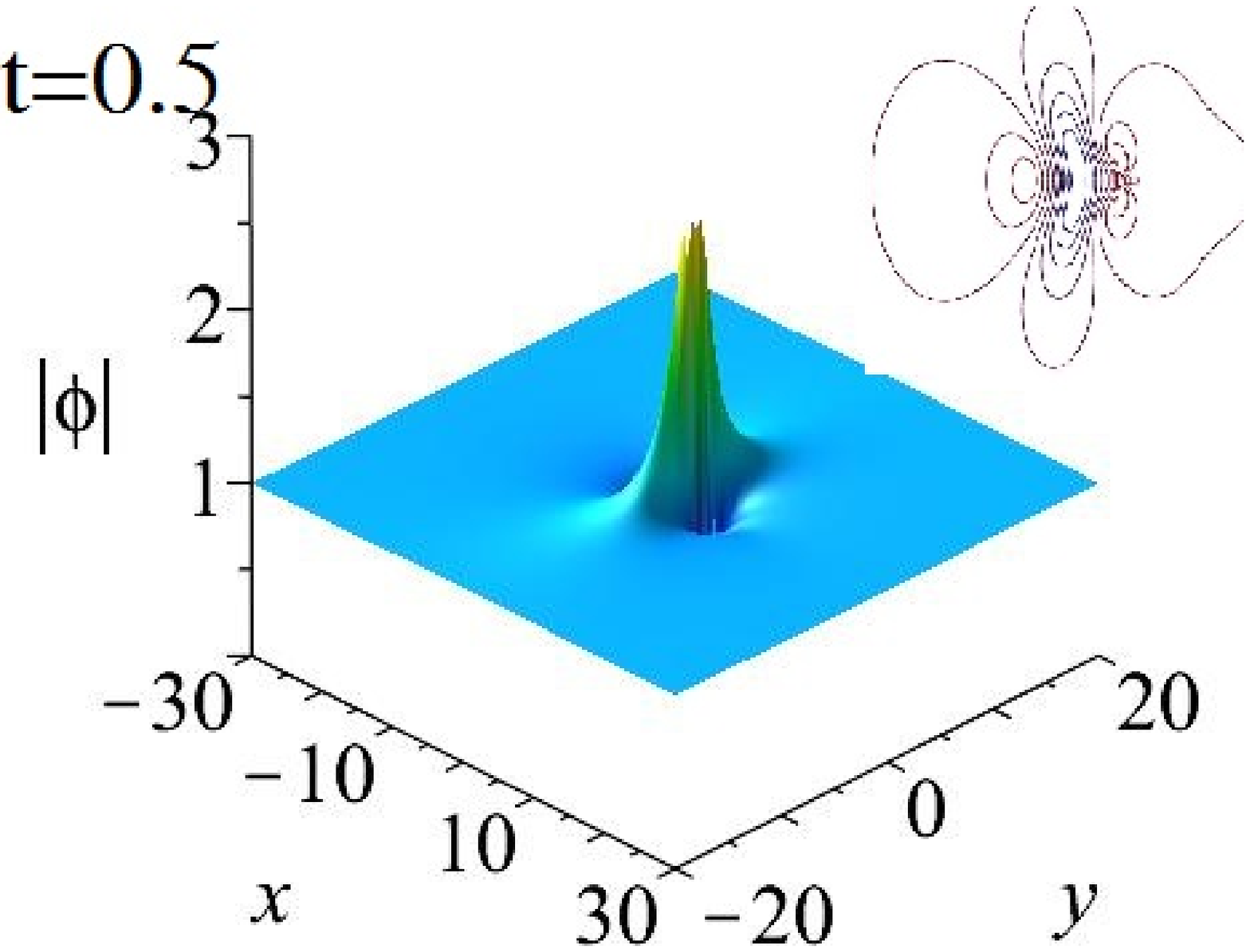}}
\subfigure{\includegraphics[width=5cm]{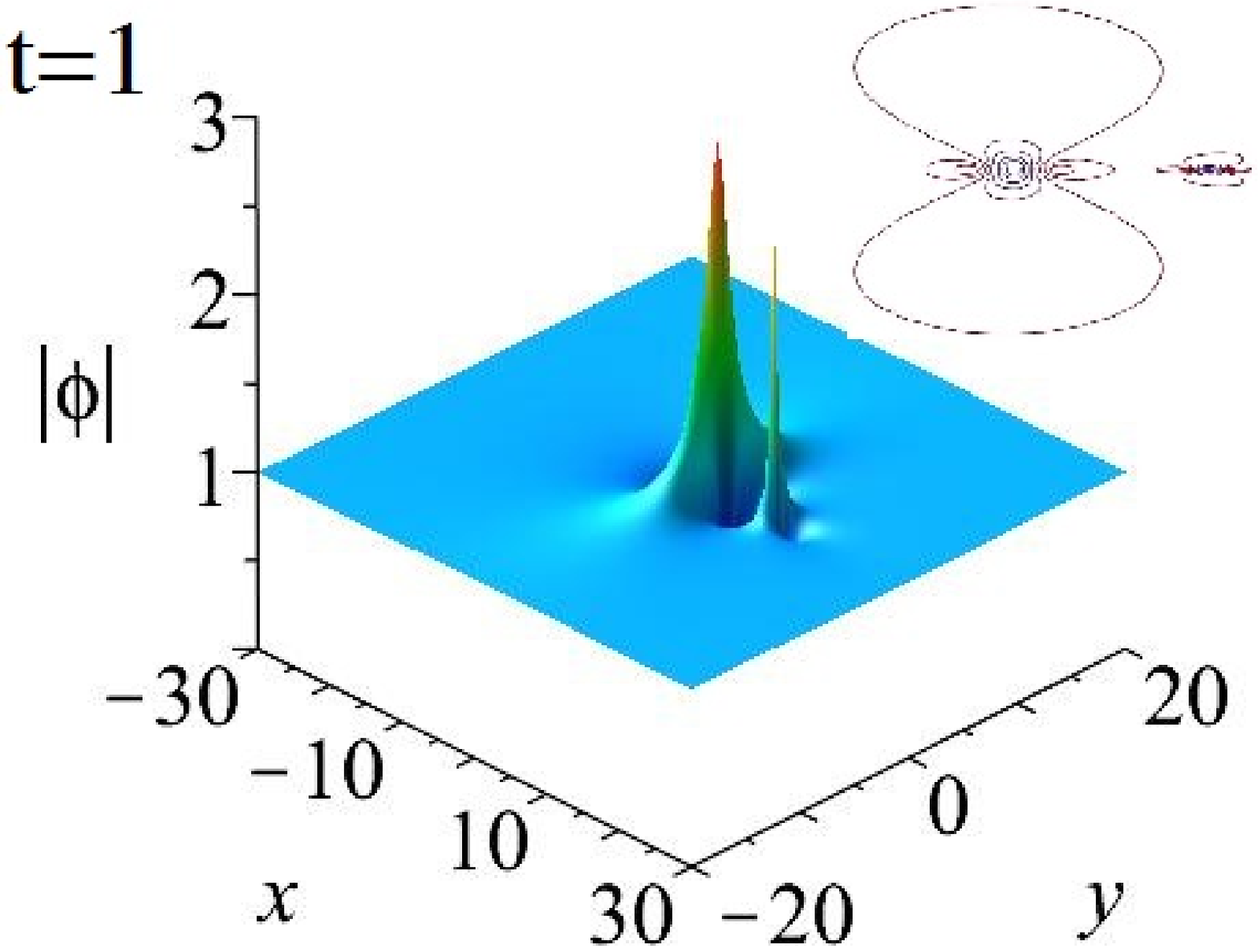}}\quad
\subfigure{\includegraphics[width=5cm]{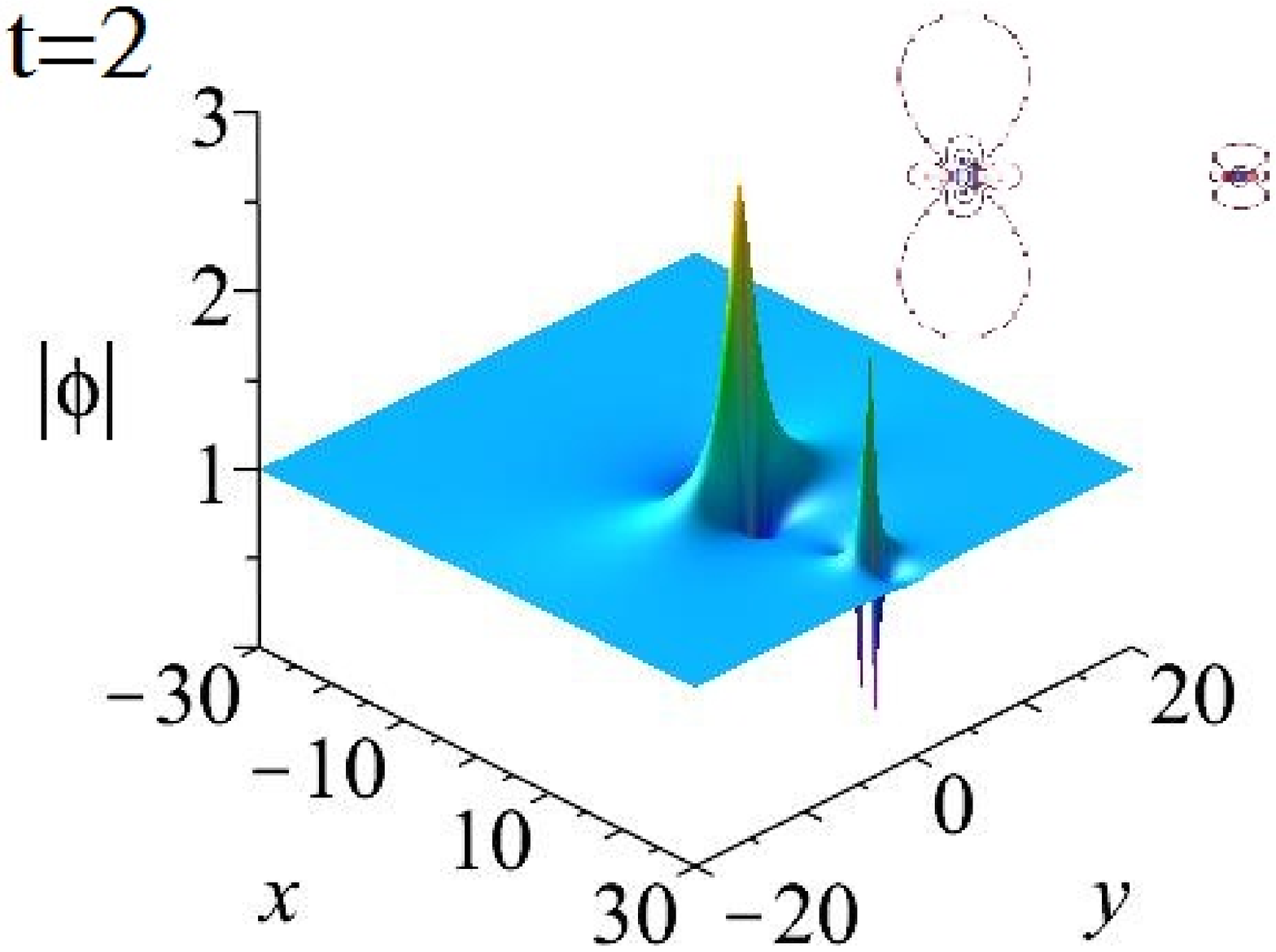}}
\caption{The superposition of two fundamental lumps at different times $t=-2,-1,0,0.5,1,2$ .~}\label{fig4}
\end{figure}
Key features of the superposition between two fundamental lumps at different times are demonstrated in
Fig.\ref{fig4}. Obviously, one lump travels at a higher speed than the other one. When they get closer, the shapes of these two lumps begin to alter. In particular, the maximum amplitude of these two lumps become lower when these two lumps immerse into each other, which do not exceed $3$ (the maximum amplitude of the fundamental lump is $3$), see the panel at $t=0.5$. That is different from the interaction of two bright lumps in the nonlocal DSI equation \cite{jiguang2}, which can generate higher amplitudes.

\begin{figure}[!htbp]
\centering
\subfigure{\includegraphics[width=5cm]{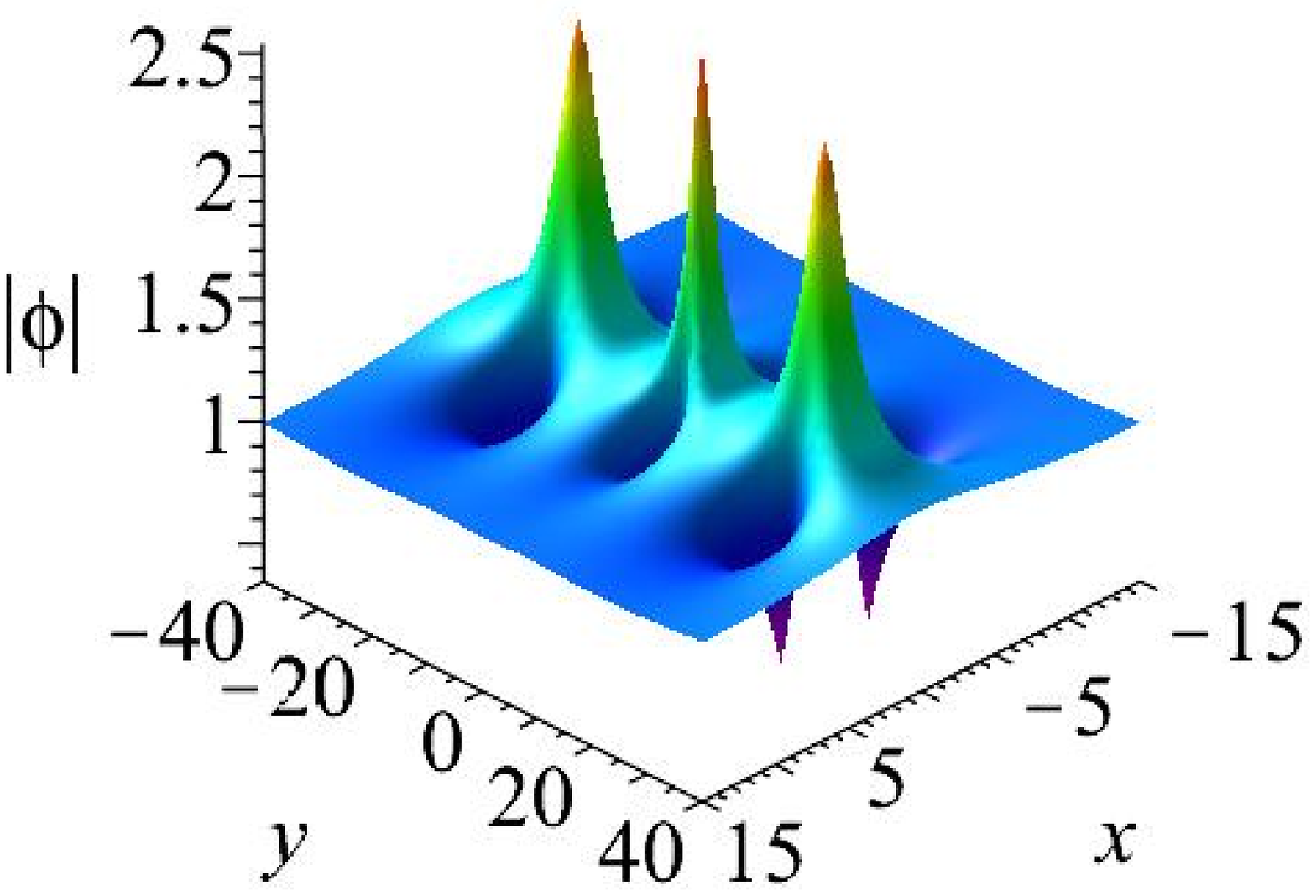}}
\subfigure{\includegraphics[width=5cm]{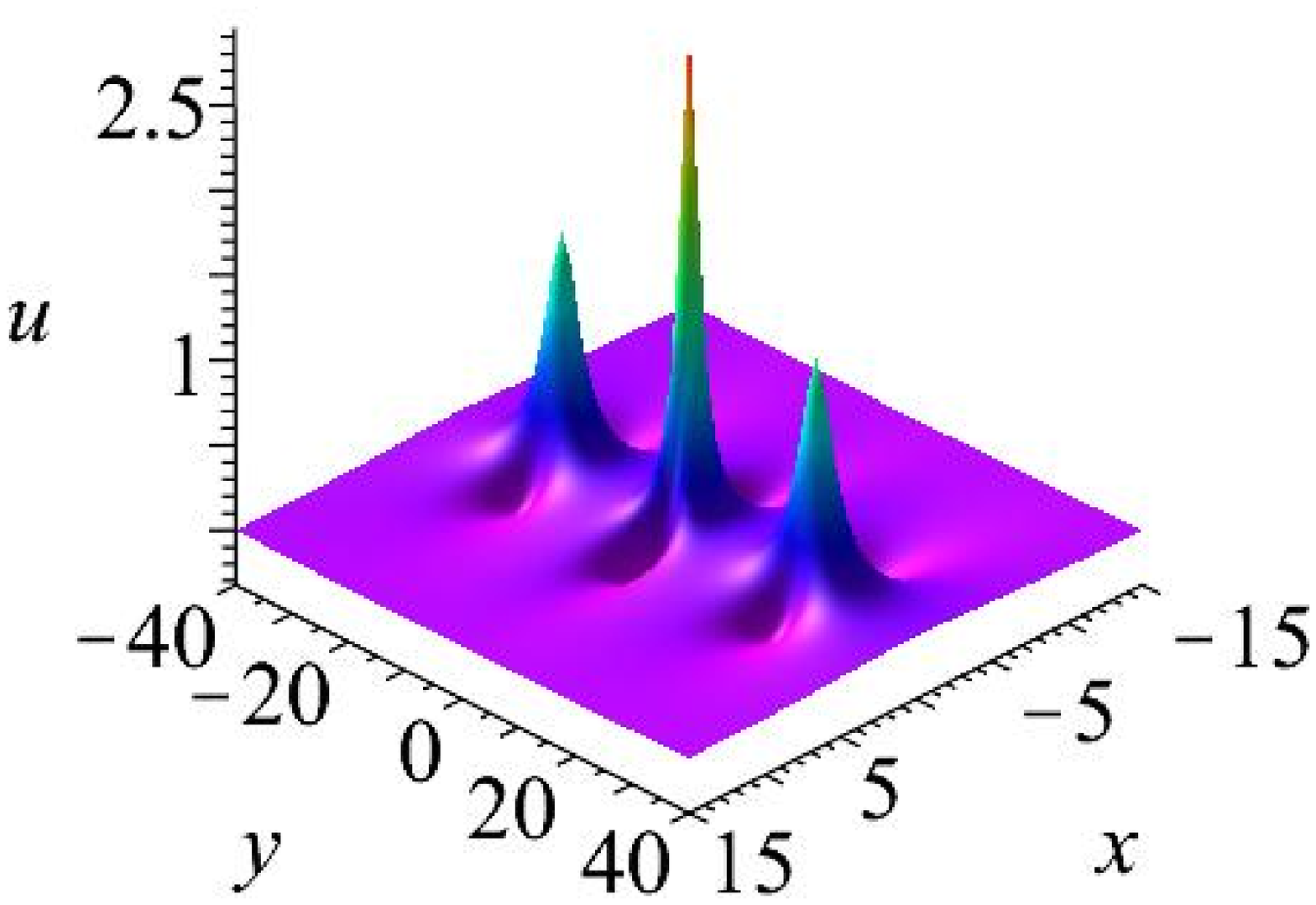}}
\caption{The third-order lump solutions $|\phi|,u$ defined in equation \eqref{3-ra}.~}\label{fig41}
\end{figure}

For larger $N$ and parameters satisfy parameters constraints \eqref{constrain}, higher-order lumps would be obtained. For instance, taking parameters in \eqref{fg}
\begin{equation} \label{xx5}
\begin{aligned}
N=6, \lambda_1=-\lambda_2=2, \lambda_3=-\lambda_4=\frac{3}{2}, \lambda_5=-\lambda_6=1,\kappa=1,
\end{aligned}
\end{equation}
the three-lump solutions can be derive, which is

\begin{equation} \label{3-ra}
\begin{aligned}
\phi=1+\frac{g_6^0}{f_6},\,u=2( {\rm log} f_{6} )_{xx},
\end{aligned}
\end{equation}
where
\begin{equation} \label{3-fg}
\begin{aligned}
f_6=&{\frac {2113125\,{t}^{4}{y}^{2}}{64}}+{\frac {61\,{x}^{4}{y}^{2}}{4}}-
{\frac {7239053785\,{t}^{3}x}{2016}}+{\frac {24747767333\,{t}^{2}{x}^{
2}}{28224}}\\
&-{\frac {551823397\,t{x}^{3}}{5880}}-{\frac {231250057\,t{y
}^{2}x}{8820}}-{\frac {425\,{y}^{4}tx}{4}}+{\frac {37515625\,{t}^{6}}{
64}}\\
&+9\,{x}^{6}+{\frac {3798316849584289\,{t}^{2}}{311169600}}+{\frac
{19715017078429\,{x}^{2}}{54022500}}+{y}^{6}+\\
&{\frac {503447329781761\,
{y}^{2}}{1944810000}}-{\frac {173731727\,{y}^{4}}{176400}}-{\frac {
314125\,{t}^{3}{y}^{2}x}{16}}-{\frac {1685\,{x}^{3}{y}^{2}t}{4}}\\
&+{
\frac {34675\,{t}^{2}{x}^{2}{y}^{2}}{8}}+{\frac {14608125\,{t}^{4}{x}^
{2}}{64}}+{\frac {90925\,{x}^{4}{t}^{2}}{16}}-{\frac {772625\,{t}^{3}{
x}^{3}}{16}}\\
&-{\frac {273805647604109\,tx}{64827000}}+{\frac {870750371
\,{t}^{2}{y}^{2}}{9408}}+{\frac {291073289\,{x}^{2}{y}^{2}}{176400}}+\\
&{
\frac {1569350425\,{t}^{4}}{288}}+{\frac {18149933\,{x}^{4}}{4900}}-{
\frac {18221875\,{t}^{5}x}{32}}-{\frac {705\,{x}^{5}t}{2}}+{\frac {
6525\,{y}^{4}{t}^{2}}{16}}\\
&+{\frac {29\,{y}^{4}{x}^{2}}{4}}+{\frac {
742769909281}{1500625}},
\end{aligned}
\end{equation}
\begin{equation} \label{3-fg1}
\begin{aligned}
g_6^{0}=A_{6}+iB_{6},
\end{aligned}
\end{equation}
and
\begin{equation} \label{3-xx}
\begin{aligned}
A_6=&x^4(-261)+x^3({
\frac {12795\,}{2}}t)+x^2(-{\frac {944525\,{t}^{2}}{16}}-{
\frac {3065\,{y}^{2}}{4}}+{\frac {16054106\,}{1225}})\\
&+x({\frac {1936375\,{t}^{3}}{8}}+9965\,t{y}^{2}-
{\frac {70472141\,t}{735}})
+(-{\frac {2970625\,{t}^{4}}{8}}-{\frac {548225\,{t}^{2}{y}^{2}}{16}}\\
&+{\frac {617733841\,{t}^{2}
}{3528}}-{\frac {1329\,{y}^{4}}{4}}+{\frac {2329695913\,{y}^{2}
}{22050}}-{\frac {3199122765724}{1500625}})    \\ \nonumber
B_6=&y[x^{4}(108)+x^{3}(-2820\,t)+x^{2}(122\,{y}^{2}+\frac {55225\,}{2}{t}^{2}-1566\,)\\
&+x(-{\frac {478625\,{t}^{3}}{4}}-1685\,t{y}^{2}-{\frac {20569361\,tx}{
1470}})+({\frac {
3093125\,{t}^{4}}{16}}+{\frac {12275\,{t}^{2}{y}^{2}}{2}}\\
&+{\frac {691962241\,{t}^{2}}{
7056}}+29\,{y}^{4}-{\frac {167205151\,{y}^{2}}{8820}}+{\frac {22387448589829
\,}{13505625}})].
\end{aligned}
\end{equation}
These third-order lump solutions  are composed of three lumps, see Fig. \ref{fig41}. But again, the maximum value of this solution $|\phi|$ stays below $3$ for all times, so this interaction does not create very high spikes either.
%%%%%%%%%%%%%%%%%%%%%%%%%%%%%%%%%%%%%

\section{Semi-rational solutions of the nonlocal Mel＊nikov equation}\label{4}
To understand resonant behaviours in the nonlocal Mel'nikov equation, we consider several types of semi-rational solutions. As derivation of the rational solutions in the last section, semi-rational solutions can also be generated by taking a long wave of the soliton solutions. Indeed, taking  a long wave limit of a part of exponential functions in $f$ and $g$ given by equation (\ref{rfg}), then this part of exponential functions is translated into rational solutions and the other part of exponential functions still keeps in exponential forms.  Hence, semi-rational solutions are generated. These semi-rational solutions describe interaction between lumps, breathers and periodic line waves. To demonstrate these unique dynamics of resonant behaviours in the nonlocal Mel'nikov equation, below we consider two cases of semi-rational solutions.

\begin{figure}[!htbp]
\centering
\subfigure{\includegraphics[width=5cm]{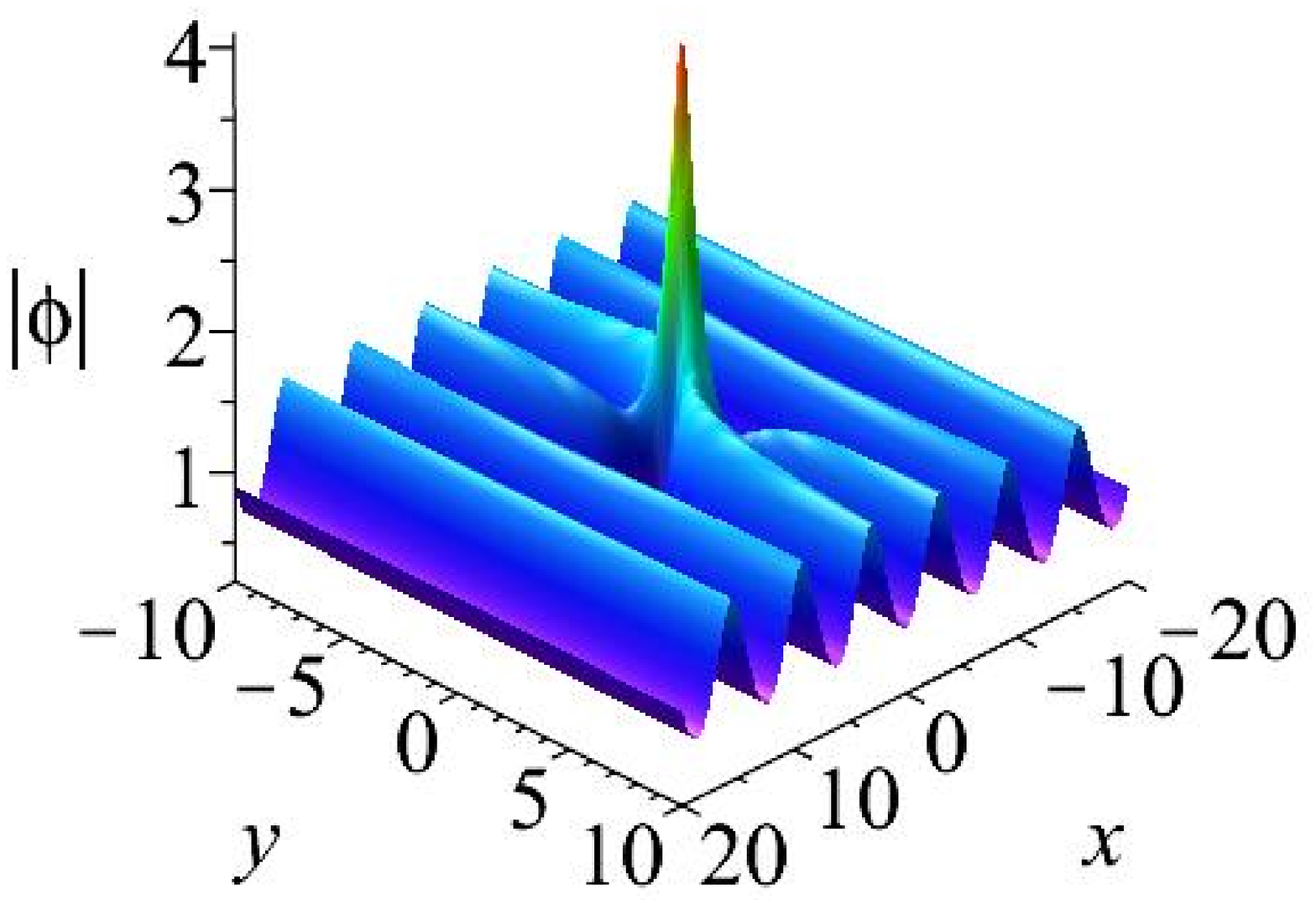}}
\subfigure{\includegraphics[width=4cm]{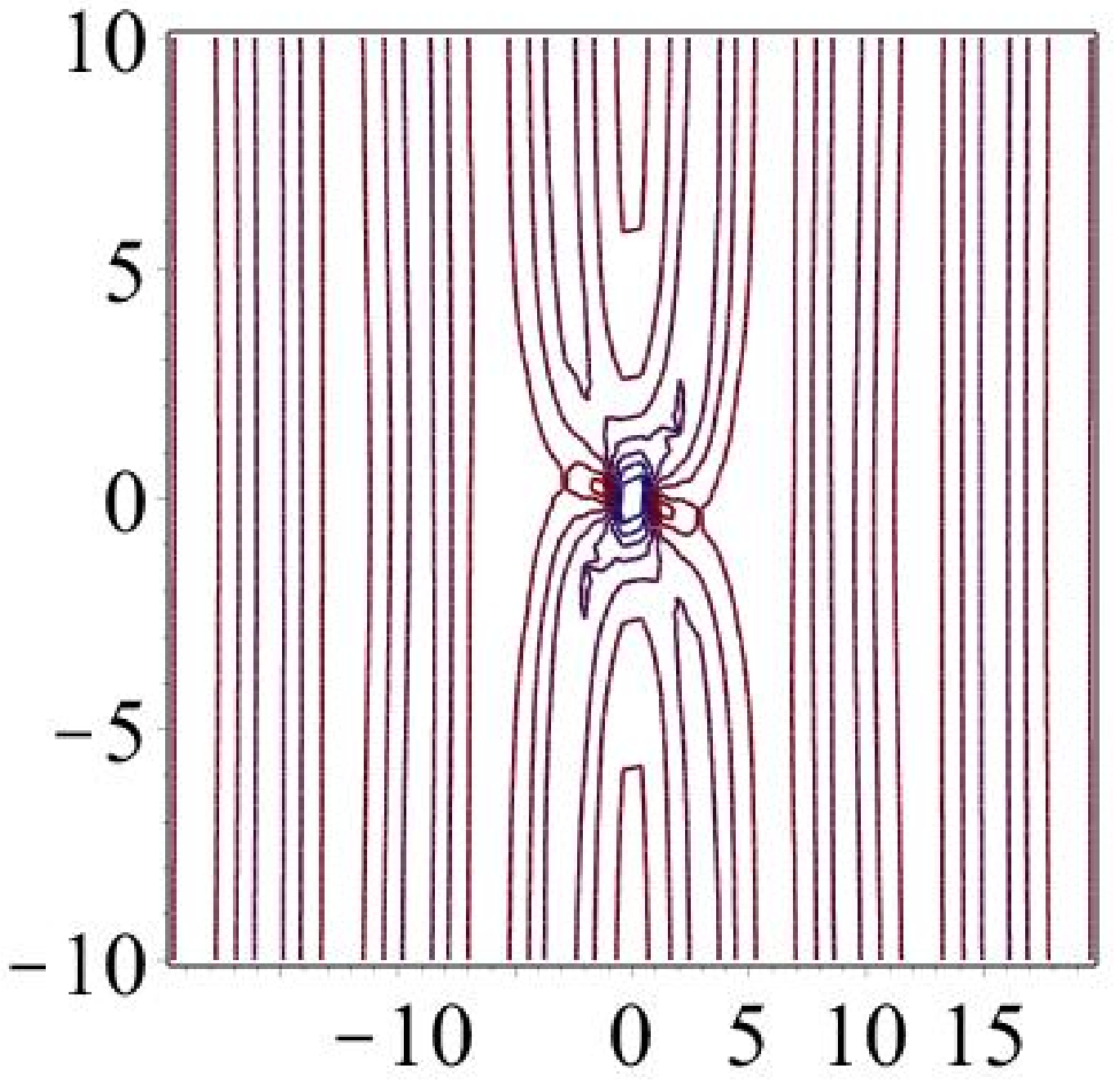}}
\subfigure{\includegraphics[width=5cm]{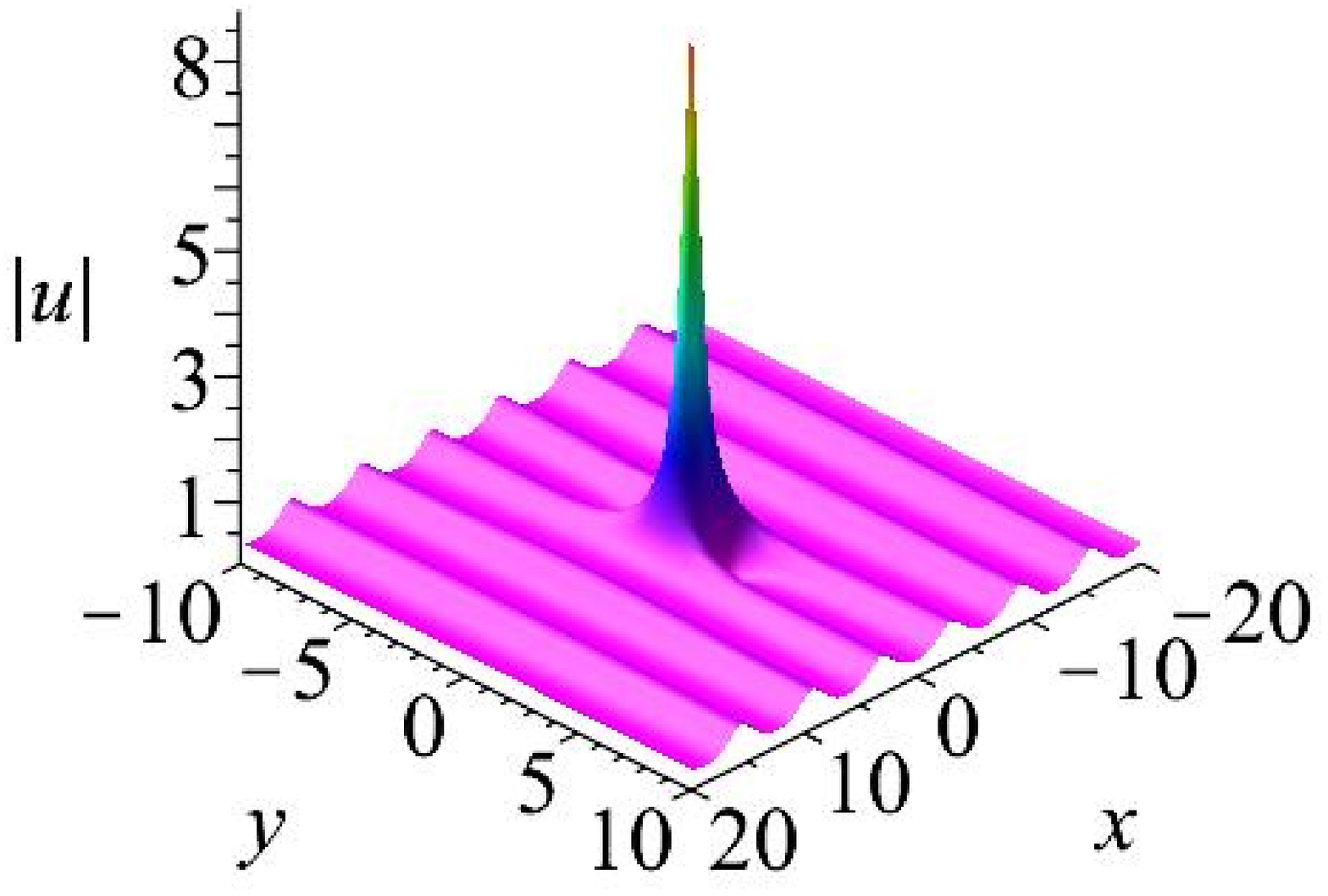}}
\subfigure{\includegraphics[width=4cm]{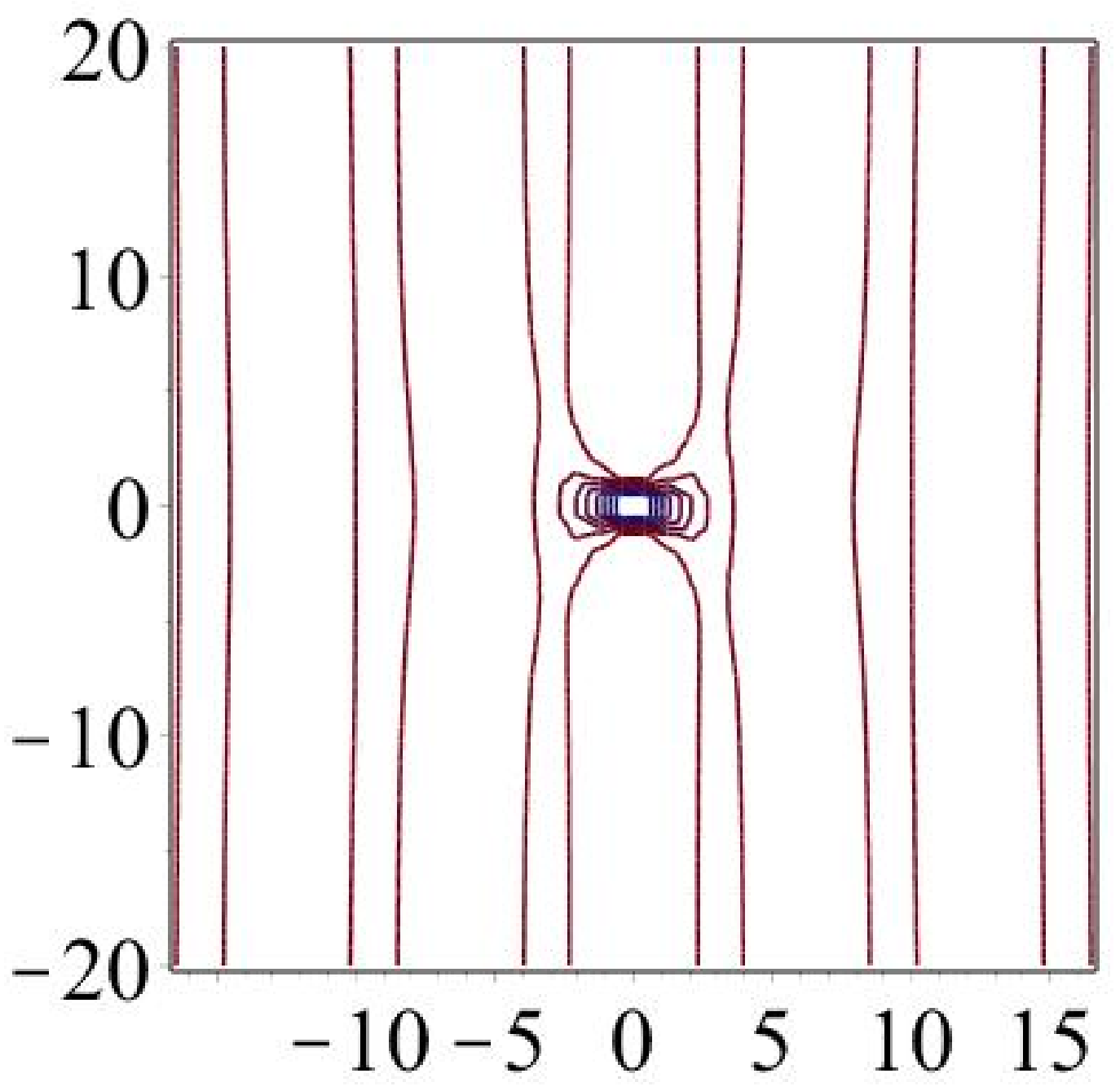}}
\caption{Semi-rational solutions constituting of a lump and periodic line waves for the local Mel'nikov equation with parameters $\lambda_0=1,P_3=1,\eta_3^0=-\frac{\pi}{6}$. The right panels are density plots of the left.~}\label{fig5}
\end{figure}

\begin{figure}[!htbp]
\centering
\subfigure{\includegraphics[width=4cm]{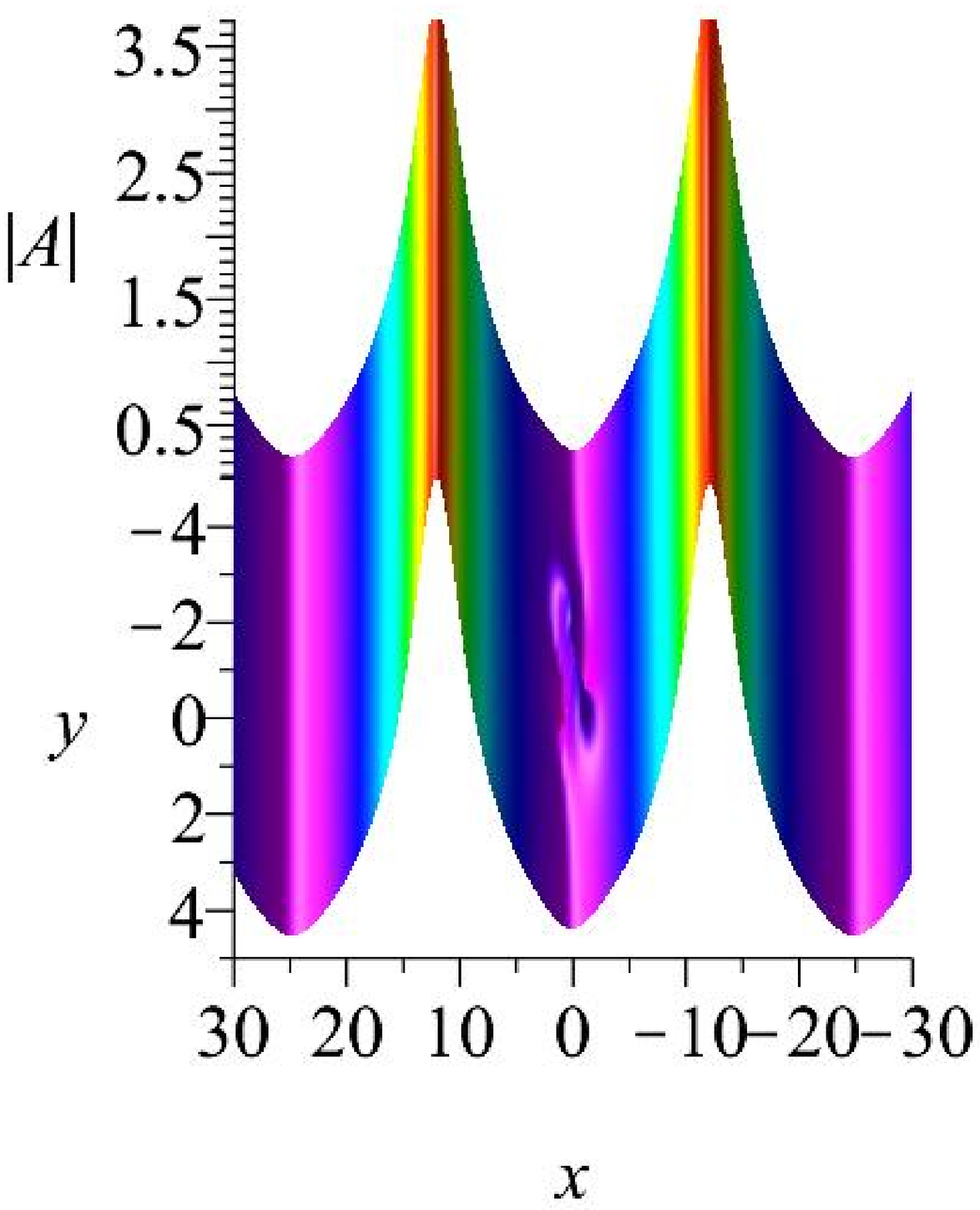}}
\subfigure{\includegraphics[width=5cm]{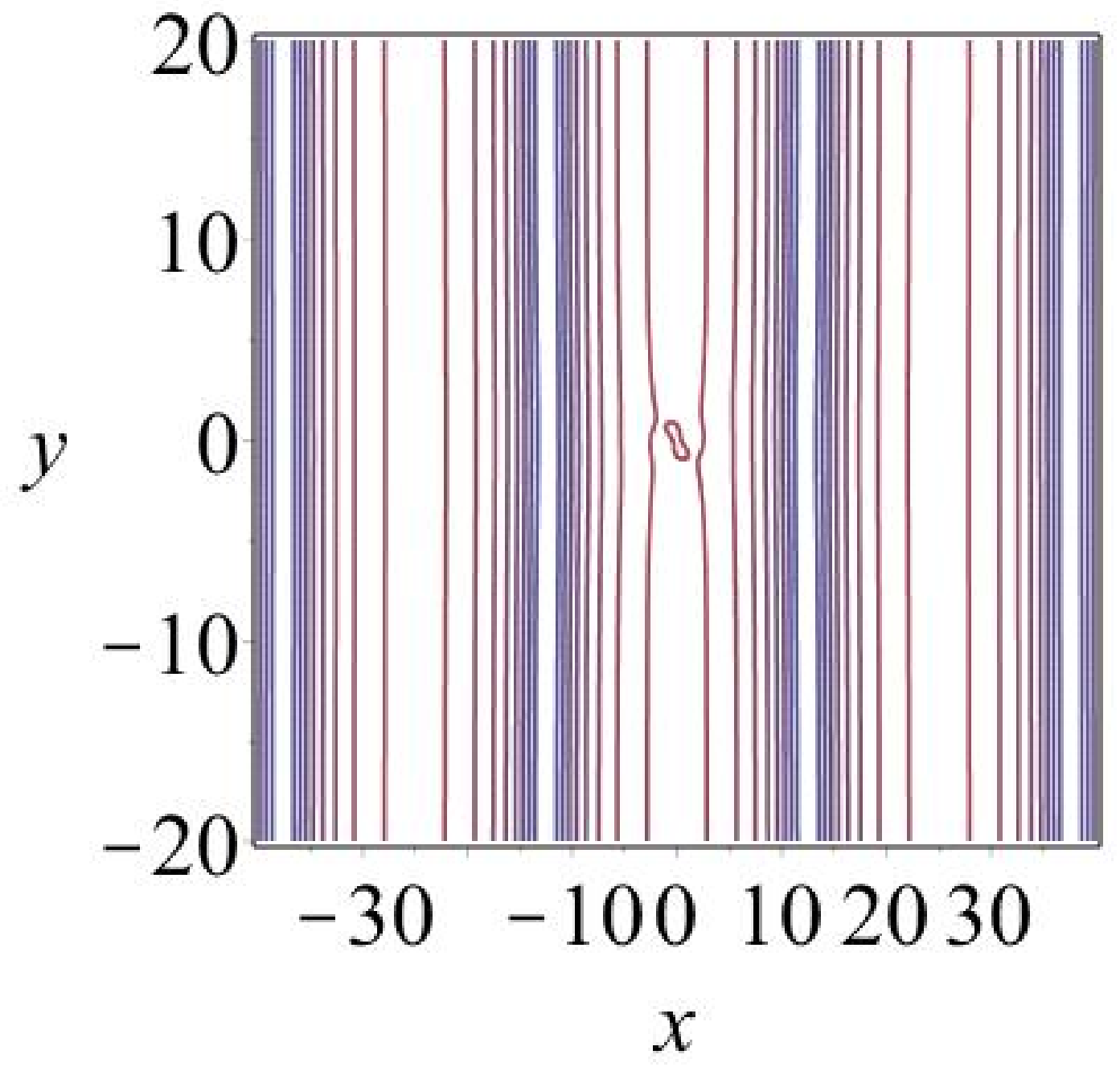}}
\subfigure{\includegraphics[width=4cm]{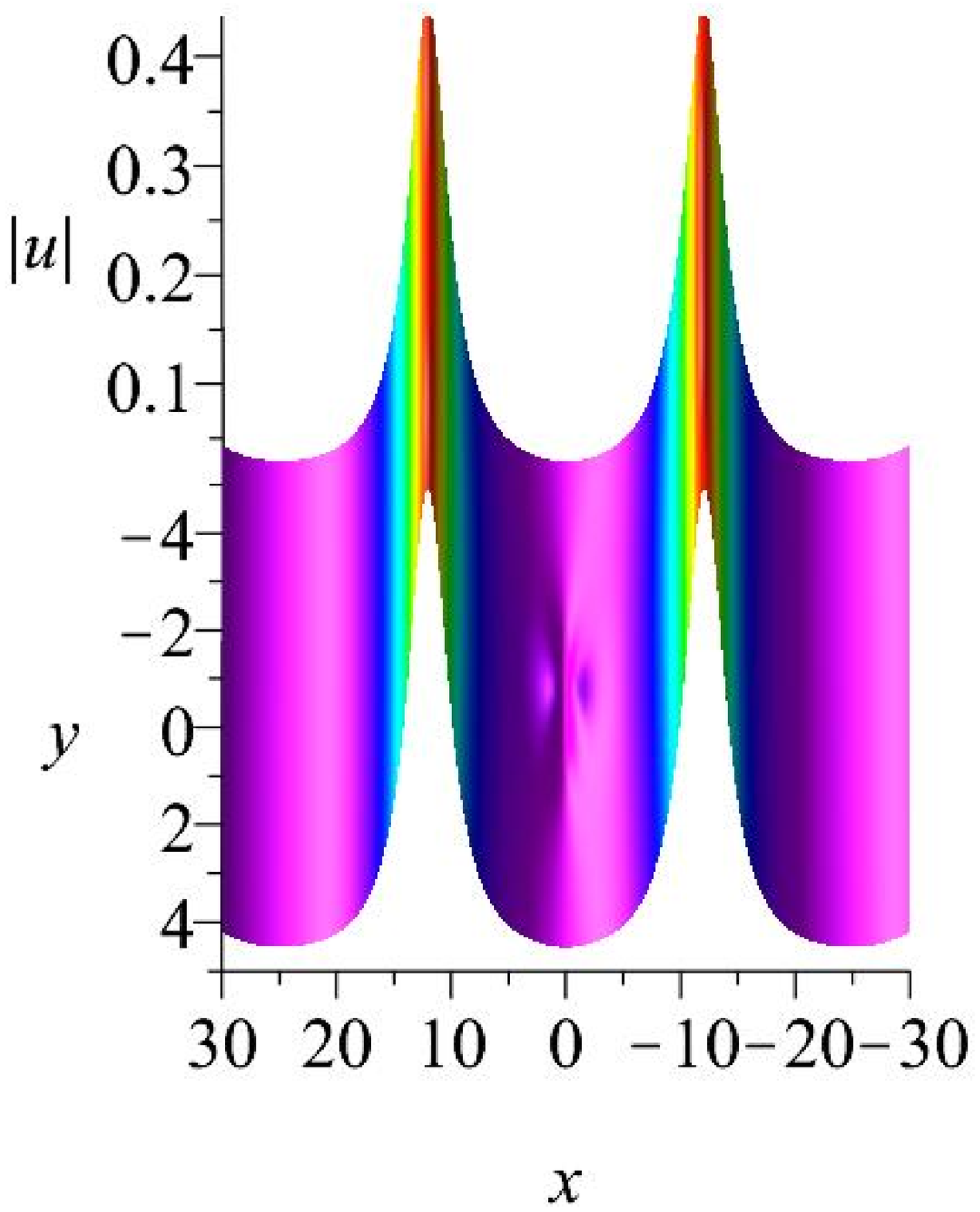}}
\subfigure{\includegraphics[width=5cm]{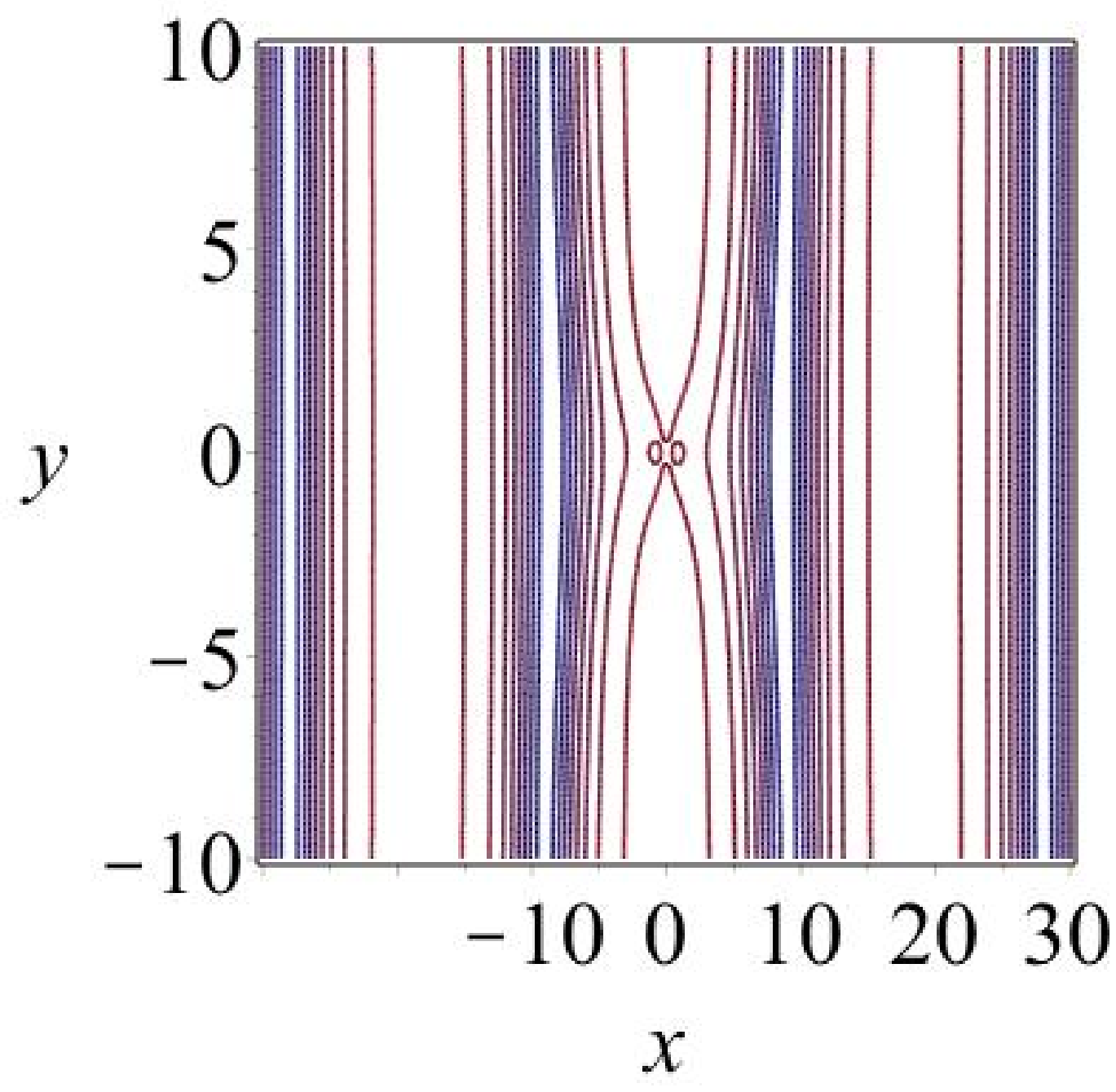}}
\caption{Semi-rational solutions constituting of a lump and periodic line waves for the local Mel'nikov equation with parameters $\lambda_0=1,P_3=1/3,\eta_3^0=-\frac{\pi}{6}$. The right panels are density plots of the left.~}\label{fig6}
\end{figure}

\noindent\textbf{Case 1: A hybrid of lumps and periodic line waves }$\\$
The simplest semi-rational solutions consisting of a lump and periodic line waves is generated from the third-order soliton solutions. Indeed,
taking parameters in (\ref{rfg})
\begin{equation} \label{pa-6}
\begin{aligned}
N=3\,,Q_{1}=\lambda_{1}P_{1}\,,Q_{2}=\lambda_{2}P_{2}\,,\exp(\eta_{1}^{0})=\exp(\eta_{2}^{0})=-1\,,Q_{3}=0,
\end{aligned}
\end{equation}
and taking a long wave limit as
$$P_{1}\,,P_{2}\rightarrow 0,$$
then functions $f$ and $g$ are rewritten as
\begin{equation}\label{hy-1}
\begin{aligned}
f=&(\theta_{1}\theta_{2}+a_{12})+(\theta_{1}\theta_{2}+a_{12}+a_{13}\theta_{2}+a_{23}\theta_{1}+a_{12}a_{23})e^{\eta_{3}}\,,\\
g=&(\theta_{1}+b_{1})(\theta_{2}+b_{2})+a_{12}+[(\theta_{1}+b_{1})(\theta_{2}+b_{2})+a_{12}\\
&+a_{13}(\theta_{2}+b_{2})+a_{23}(\theta_{1}+b_{1})+a_{12}a_{23}]e^{\eta_{3}+i\phi_{3}}\,,
\end{aligned}
\end{equation}
where $$a_{s3}=-\frac{4P_3^2}{P_3^2+\lambda_s^2}$$  and $a_{12}\,,b_{s}\,,\phi_{s},\eta_{3}$ are given by \eqref{rt} and \eqref{cs1}. Further,
taking parameter constraints
\begin{equation} \label{py-1}
\begin{aligned}
\lambda_{1}=-\lambda_{2}=\lambda_{0}\,,
\end{aligned}
\end{equation}
then corresponding semi-rational solutions  consisting of a lump and periodic line waves are obtained, see Figs.\ref{fig5},\ref{fig6}.  It is seen that, the periodic line waves coexisting with a fundamental lump are periodic in $x$ direction and localized in $y$ direction and the period is $\frac{2\pi}{P_{3}}$. By comparing to this type of semi-rational solutions of the nonlocal DSI equation in Ref. \cite{jiguang2}, a different phenomenon is that, the interaction between the lump and periodic line waves can generated either much higher peaks or lower peaks.  In Fig.\ref{fig5}, the maximum amplitude of the lump can reach $4$ (four times the constant background), while it is higher than the fundamental lumps. However, in the Fig. \ref{fig6},  the maximum amplitudes of the lump do not exceed $2$ (two times the constant background). In this case, the interaction between the lump and periodic line waves generate lower peaks. Note that this type of semi-rational solutions has not been discussed in the local Mel'nikov equation before.

 In particular, when one takes
\begin{equation}\label{xx3}
\begin{aligned}
\lambda=\frac{\sqrt{6}}{3},\kappa=-1, \Omega_3=P_3
\end{aligned}
\end{equation}
in equation \eqref{hy-1}, and then taking the variable transformations defined in equation \eqref{MStr}, the corresponding mixed solutions \eqref{hy-1} reduce to one dimensional mixed solutions consisting of a fundamental rogue wave and periodic line waves to the nonlocal Schr\"odinger-Boussinesq equation.

\begin{figure}[!htbp]
\centering
\subfigure{\includegraphics[width=5cm]{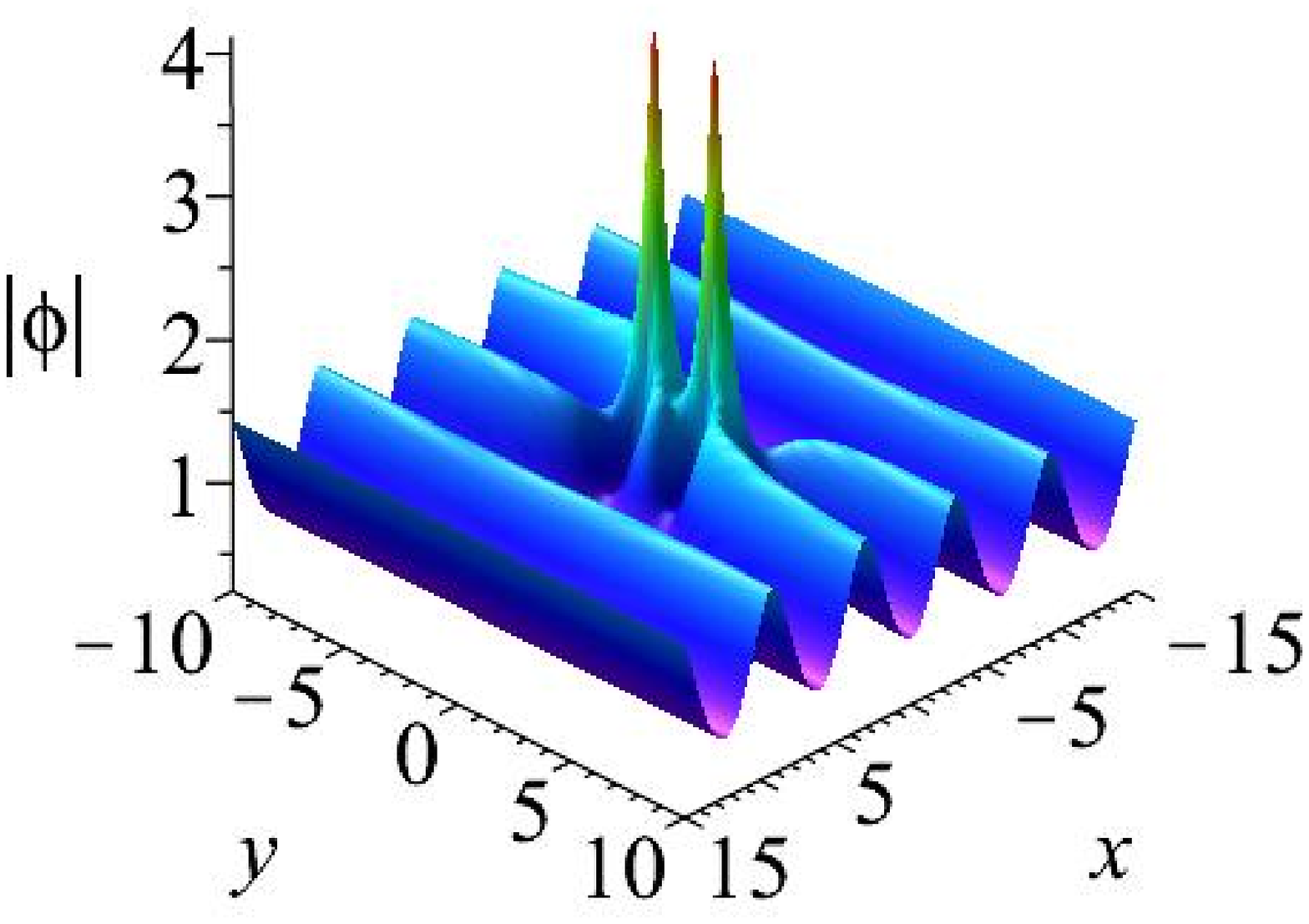}}
\subfigure{\includegraphics[width=4cm]{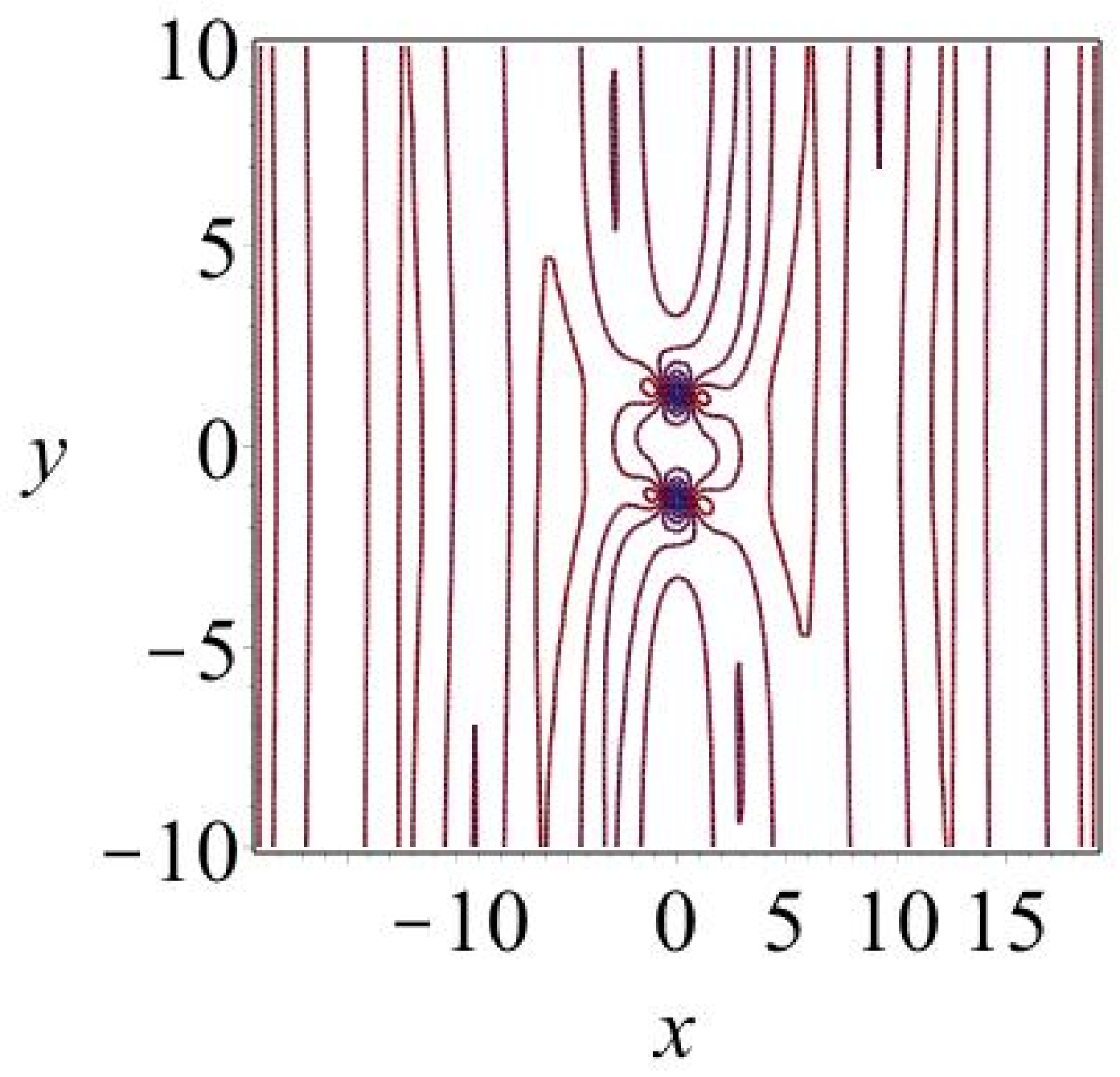}}
\subfigure{\includegraphics[width=5cm]{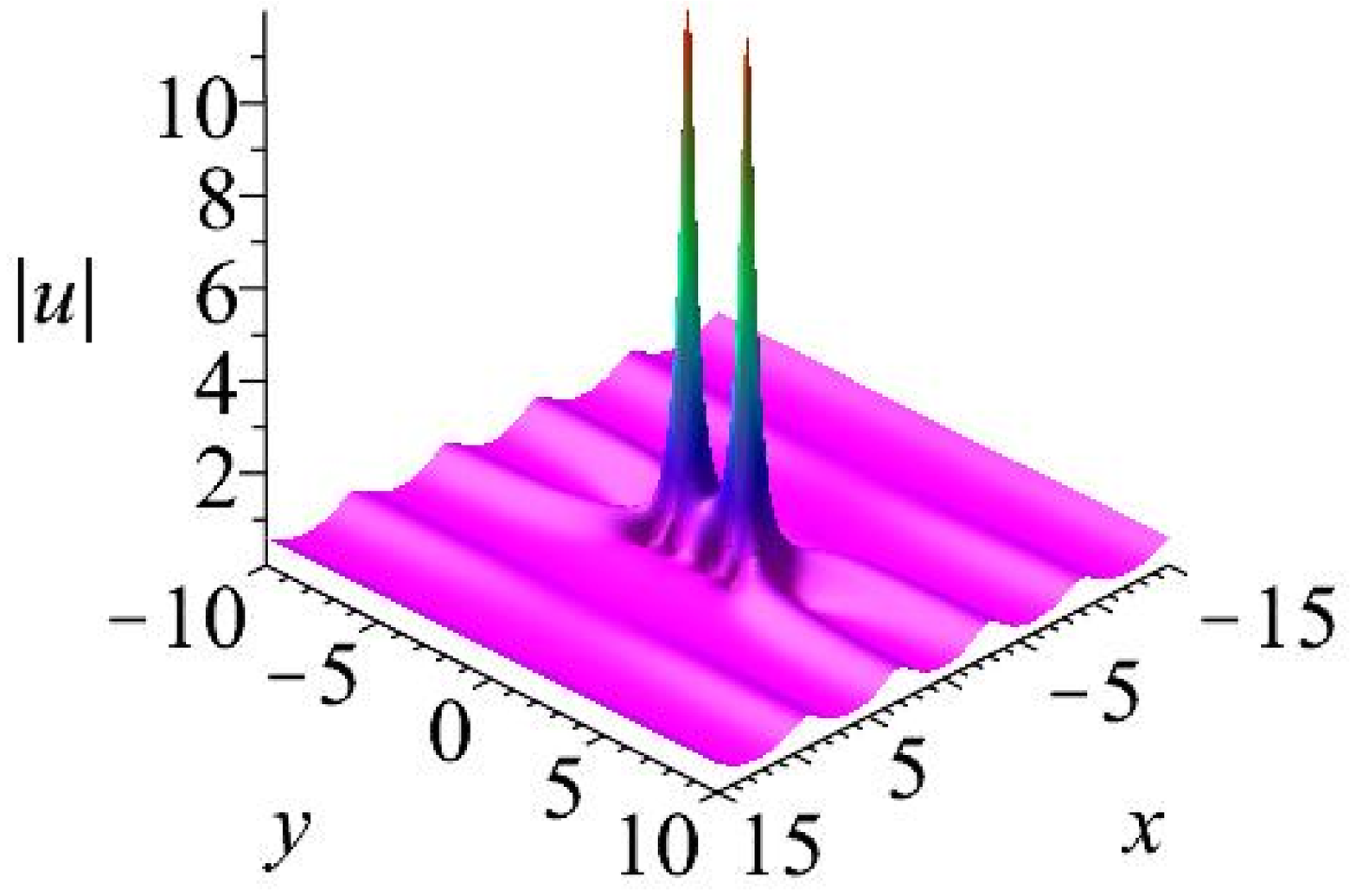}}
\subfigure{\includegraphics[width=4cm]{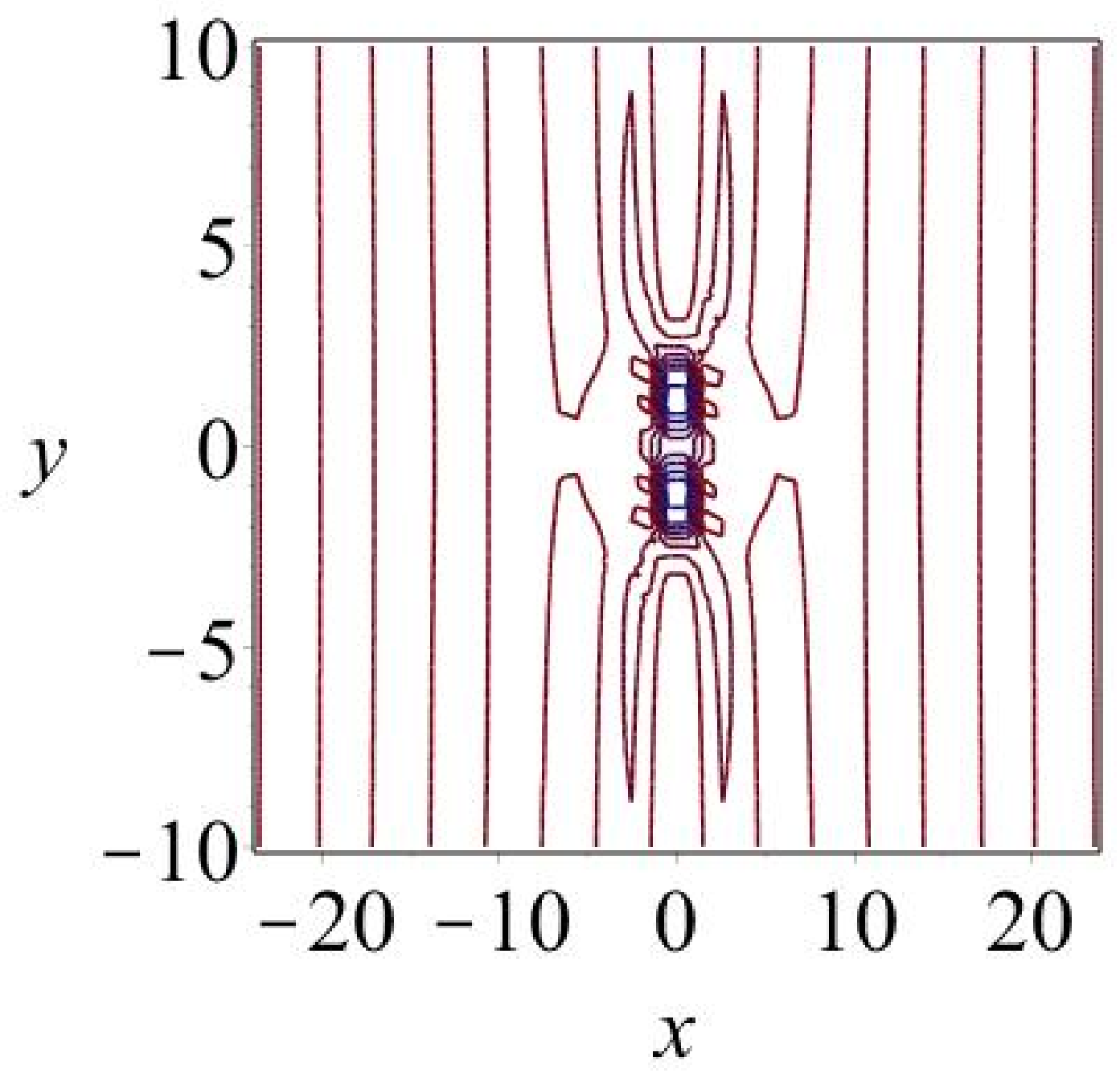}}
\caption{Semi-rational solutions constituting of two lump and periodic line waves for the local Mel'nikov equation with parameters $\kappa=1,\lambda_1=-\lambda_2=1,\lambda_3=-\lambda_4=2, P_5=1,Q_5=0,\eta_5^0=-\frac{\pi}{6}$. The right panels are density plots of the left.~}\label{fig9}
\end{figure}

Semi-rational solutions consisting of more lumps and periodic line waves can also be generated by a similar way for larger $N$. Below we consider a subclass of semi-rational solutions composed of two lumps and periodic line waves,  which can be generated from $5$-soliton solutions. Indeed,
setting in parameters in equation \eqref{rfg}
\begin{equation} \label{pa-6}
\begin{aligned}
N=5\,,Q_{j}=\lambda_{j}P_{j}\,,\exp(\eta_{j}^{0})=-1\,,Q_{5}=0\,(j=1,2,3,4)\,,
\end{aligned}
\end{equation}
and then taking a long wave limit as
$$P_{j}\rightarrow 0\,(j=1,2,3,4),$$
then functions $f$ and $g$ are translated from  exponential functions to a combination of rational and exponential functions
%{\scriptsize}
\begin{equation} \label{ff3}
\begin{aligned}
f=&(\theta_{1}\theta_{2}\theta_{3}\theta_{4}+a_{12}\theta_{3}\theta_{4}+a_{13}\theta_{2}\theta_{4}+a_{14}\theta_{2}\theta_{3}+a_{23}\theta_{1}\theta_{4}
+a_{24}\theta_{1}\theta_{3}+a_{34}\theta_{1}\theta_{2}\\
&+a_{12}a_{34}+
a_{13}a_{24}+a_{14}a_{23})+e^{\eta_{5}}[\theta_{1}\,\theta_{2}\,\theta_{3}\,\theta_{4}+a_{45}\,\theta_{1}\,\theta_{2}\,\theta_{3}+a_{35}\,\theta_{1}\,\theta_{2}\,\theta_{4}
\\&+a_{25}\,\theta_{1}\,\theta_{3}\,\theta_{4}+a_{15}\theta_{2}\,\theta_{3}\,\theta_{4}+
(a_{35}a_{45}+a_{34})\theta_{1}\,\theta_{2}+(a_{25} a_{45}+a_{24})\,\theta_{1}\,\theta_{3}+\\
&(a_{25}a_{35}+a_{23})\,\theta_{1}\,\theta_{4}+(a_{15}a_{45}+a_{14})\,\theta_{2}\,\theta_{3}+(a_{15}a_{35}+a_{13})\,\theta_{2}\,\theta_{4}+(a_{15}a_{25}\\
&+a_{12})\theta_{3}\,\theta_{4}+(a_{25}a_{35}\,a_{45}+a_{23}a_{45}+a_{25}a_{34}+a_{24}a_{35})\,\theta_{1}+(a_{15}a_{35}a_{45}+\\
&a_{14}a_{35}+a_{13}a_{45}+a_{15}a_{34})\,\theta_{2}+(a_{15} a_{25}a_{45}+a_{14}a_{25}+a_{15}a_{24}+a_{12}a_{45})\,\theta_{3}\\
&+(a_{15}a_{25}a_{35}+a_{15}a_{23}+a_{13}a_{25}+a_{12}a_{35})\,\theta_{4}+a_{12}a_{34}+a_{13}\,a_{24}+a_{14}a_{23}\\
&+a_{12}\,a_{35}\,a_{45}+a_{13}a_{25}a_{45}+a_{14}a_{25}a_{35}+a_{15}a_{24}a_{35}+a_{15}a_{25}a_{34}\,+a_{15}a_{23}a_{45}\\
&+a_{15}a_{25} a_{35} a_{45}]\,,\\
g=&[(\theta_{1}+b_{1})(\theta_{2}+b_{2})(\theta_{3}+b_{3})(\theta_{4}+b_{4})+a_{12}(\theta_{3}+b_{3})(\theta_{4}+b_{4})+a_{13}(\theta_{2}+b_{2})(\theta_{4}\\
&+b_{4})+a_{14}(\theta_{2}+b_{2})(\theta_{3}+b_{3})+a_{23}(\theta_{1}+b_{1})(\theta_{4}+b_{4})
+a_{24}(\theta_{1}+b_{1})(\theta_{3}+b_{3})\\
&+a_{34}(\theta_{1}+b_{1})(\theta_{2}+b_{2})+a_{12}a_{34}+a_{13}a_{24}+a_{14}a_{23}]+e^{\eta_{5}+i\phi_{5}}[(\theta_{1}+b_{1})(\theta_{2}+\\
&b_{2})(\theta_{3}+b_{3})(\theta_{4}+b_{4})+a_{45}\,(\theta_{1}+b_{1})(\theta_{2}+b_{2})(\theta_{3}+b_{3})+a_{35}\,(\theta_{1}+b_{1})(\theta_{2}+b_{2})\\
&(\theta_{4}+b_{4})
+a_{25}\,(\theta_{1}+b_{1})(\theta_{3}+b_{3})(\theta_{4}+b_{4})+a_{15}(\theta_{2}+b_{2})(\theta_{3}+b_{3})(\theta_{4}+b_{4})+
\\
&(a_{35}a_{45}+a_{34})(\theta_{1}+b_{1})(\theta_{2}+b_{2})+(a_{25} a_{45}+a_{24})\,(\theta_{1}+b_{1})(\theta_{3}+b_{3})+(a_{25}a_{35}\\
&+a_{23})(\theta_{2}+b_{2})(\theta_{4}+b_{4})+(a_{15}a_{45}+a_{14})(\theta_{2}+b_{2})(\theta_{3}+b_{3})+(a_{15}a_{35}+a_{13})\,\\
&(\theta_{2}+b_{2})(\theta_{4}+b_{4})+(a_{15}a_{25}+a_{12})(\theta_{3}+b_{3})(\theta_{4}+b_{4})+(a_{25}a_{35}\,a_{45}+a_{23}a_{45}\\
&(\theta_{2}+b_{2})+(a_{15} a_{25}a_{45}+a_{14}a_{25}+a_{15}a_{24}+a_{12}a_{45})\,(\theta_{3}+b_{3})+(a_{15}a_{25}a_{35}+\\
&a_{15}a_{23}+a_{13}a_{25}+a_{12}a_{35})(\theta_{4}+b_{4})+a_{12}(a_{34}+a_{35}\,a_{45})+a_{13}(\,a_{24}+a_{25}a_{45})\\
&+a_{14}(a_{23}+a_{25}a_{35})+a_{15}(a_{24}a_{35}+a_{25}a_{34}\,+a_{23}a_{45}+a_{25} a_{35} a_{45})],
\end{aligned}
\end{equation}
where $$a_{j5}=-\frac{4P_5^2}{P_5^2+\lambda_j^2},$$ and $\theta_{j}\,,b_{j}\,,a_{ij}\,(\,1\leq i<j\leq4)$ are given by \eqref{rt}, $\eta_{5}\,,\phi_{5}$ are given by  \eqref{cs1}. Under parameter constraints
\begin{equation} \label{pa-7}
\begin{aligned}
\lambda_{1}=-\lambda_{2},\lambda_{3}=-\lambda_{4},
\end{aligned}
\end{equation}
then a family of semi-rational solutions describing two lumps on a  background of periodic line waves is generated.
These two solutions $\phi$ and $u$ with parameters
\begin{equation} \label{pa-8}
\begin{aligned}
\lambda_{1}=-\lambda_{2}=1,\lambda_{3}=-\lambda_{4}=4, P_5=1
\end{aligned}
\end{equation}
are shown in Fig. \ref{fig9}. It is seen that these two solutions $u$ and $\phi$ are composed of two lumps and periodic line waves. The period of these line waves is $2\pi$. The maximum value of the solution $\phi$ can exceed $4.5$ ($4.5$ times the constant background) at $t=0$, while the  $|u|$ can reach $12$. So this interaction between these two lumps and periodic line waves can create very high spikes either under proper parameter choices.

\begin{figure}[!htbp]
\centering
\subfigure{\includegraphics[width=5.5cm]{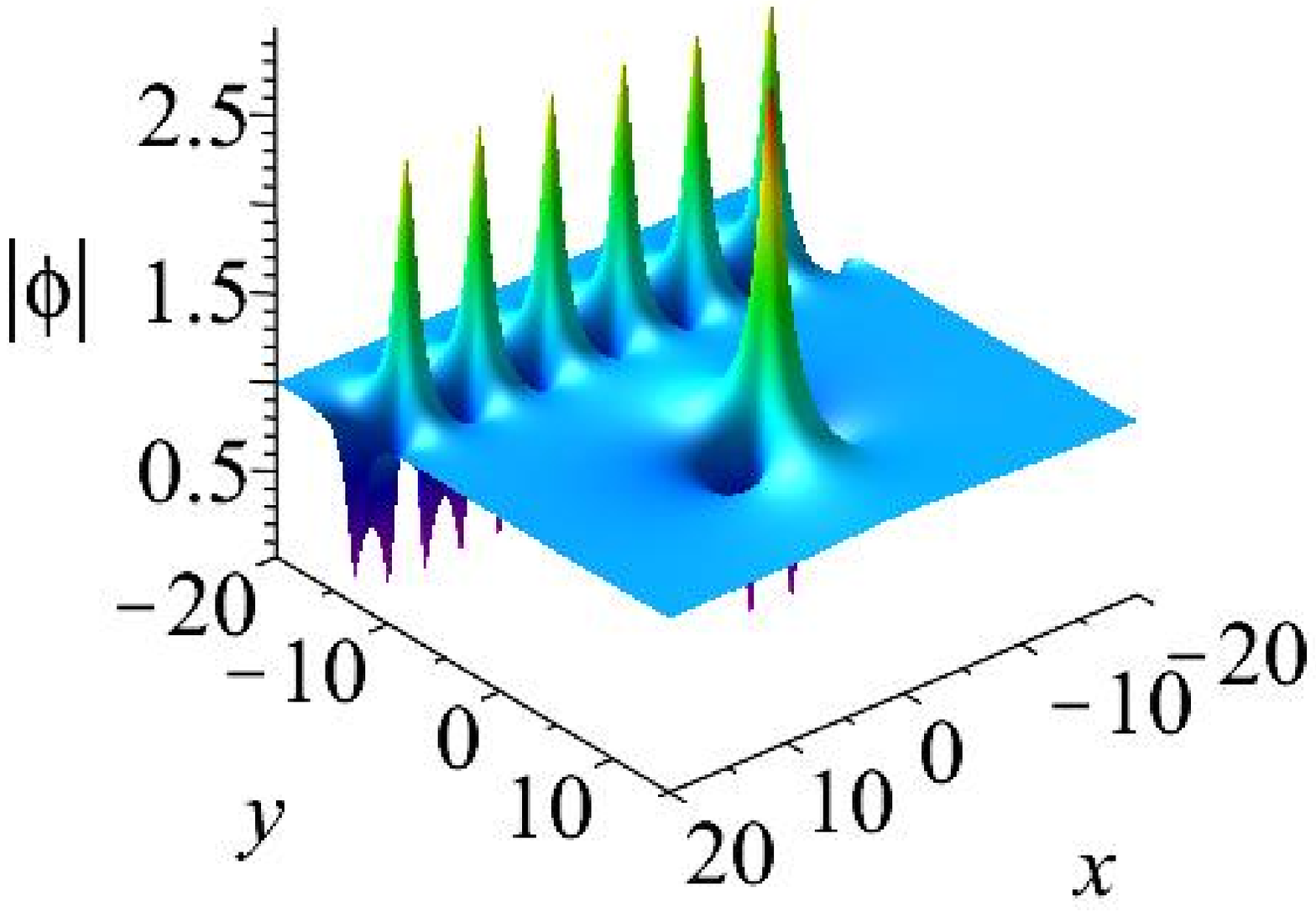}}
\subfigure{\includegraphics[width=5.5cm]{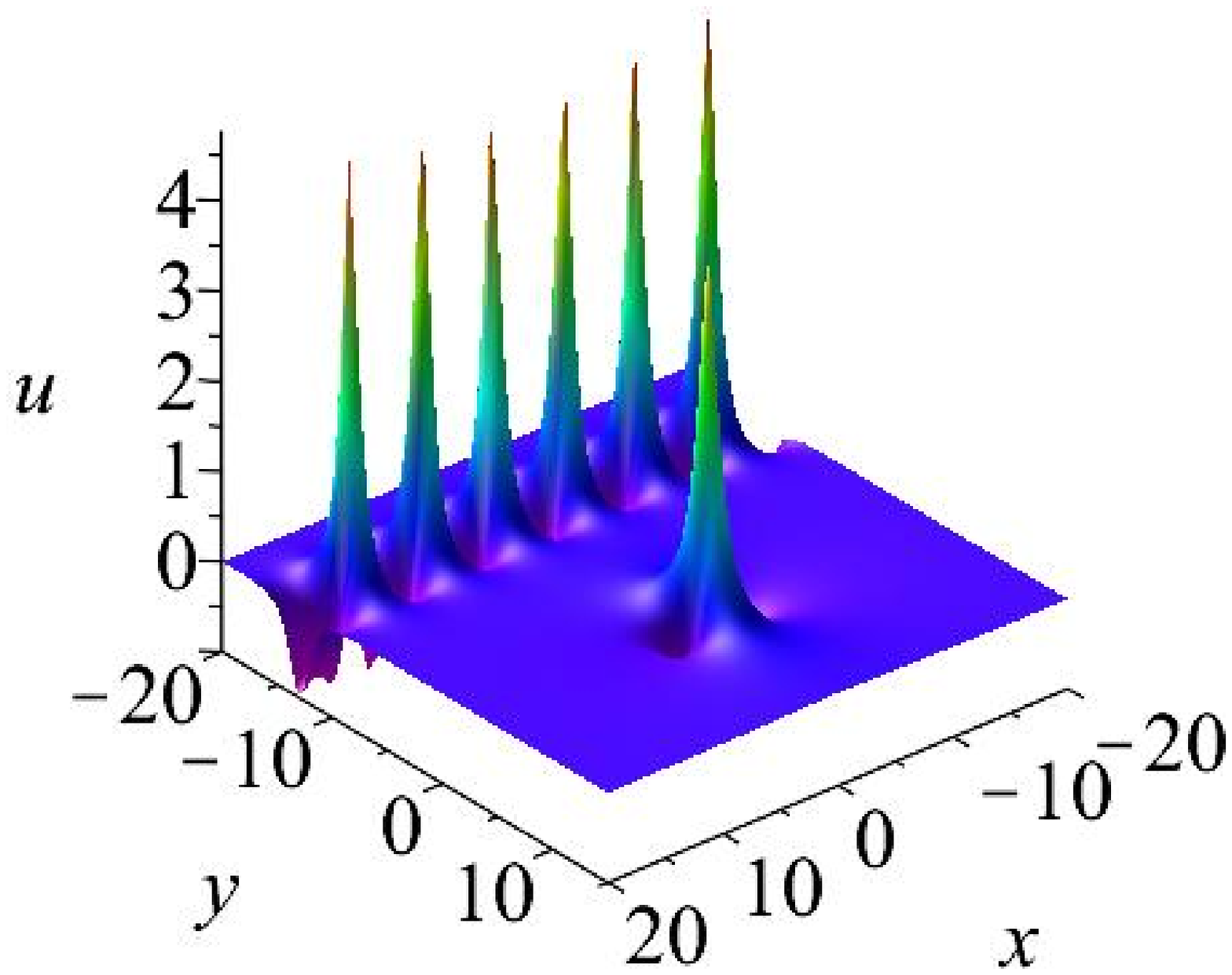}}
\caption{Semi-rational solutions $|\phi|,u$ constituting of a lump and one-breather for the local Mel'nikov equation with parameters $\kappa=1,\lambda_1=1,\lambda2=-1,P_3=-P_4=1, Q_3=Q_4=1, \eta_3^0=\eta_4^0=-2\pi,t=0$.~}\label{fig10}
\end{figure}

\begin{figure}[!htbp]
\centering
\subfigure[$\eta_{3}^0=-2\pi$]{\includegraphics[width=5cm]{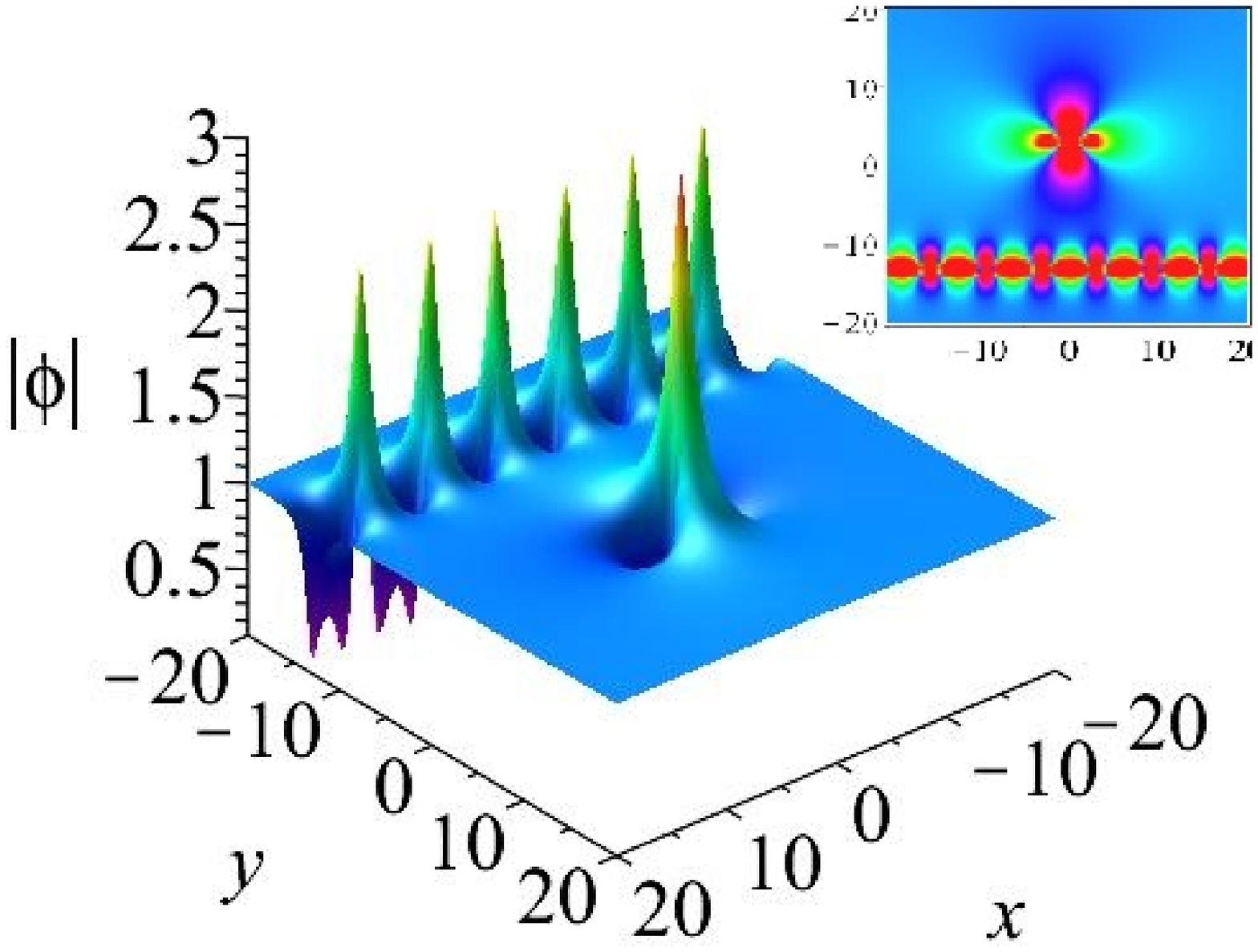}}
\subfigure[$\eta_{3}^0=0$]{\includegraphics[width=5cm]{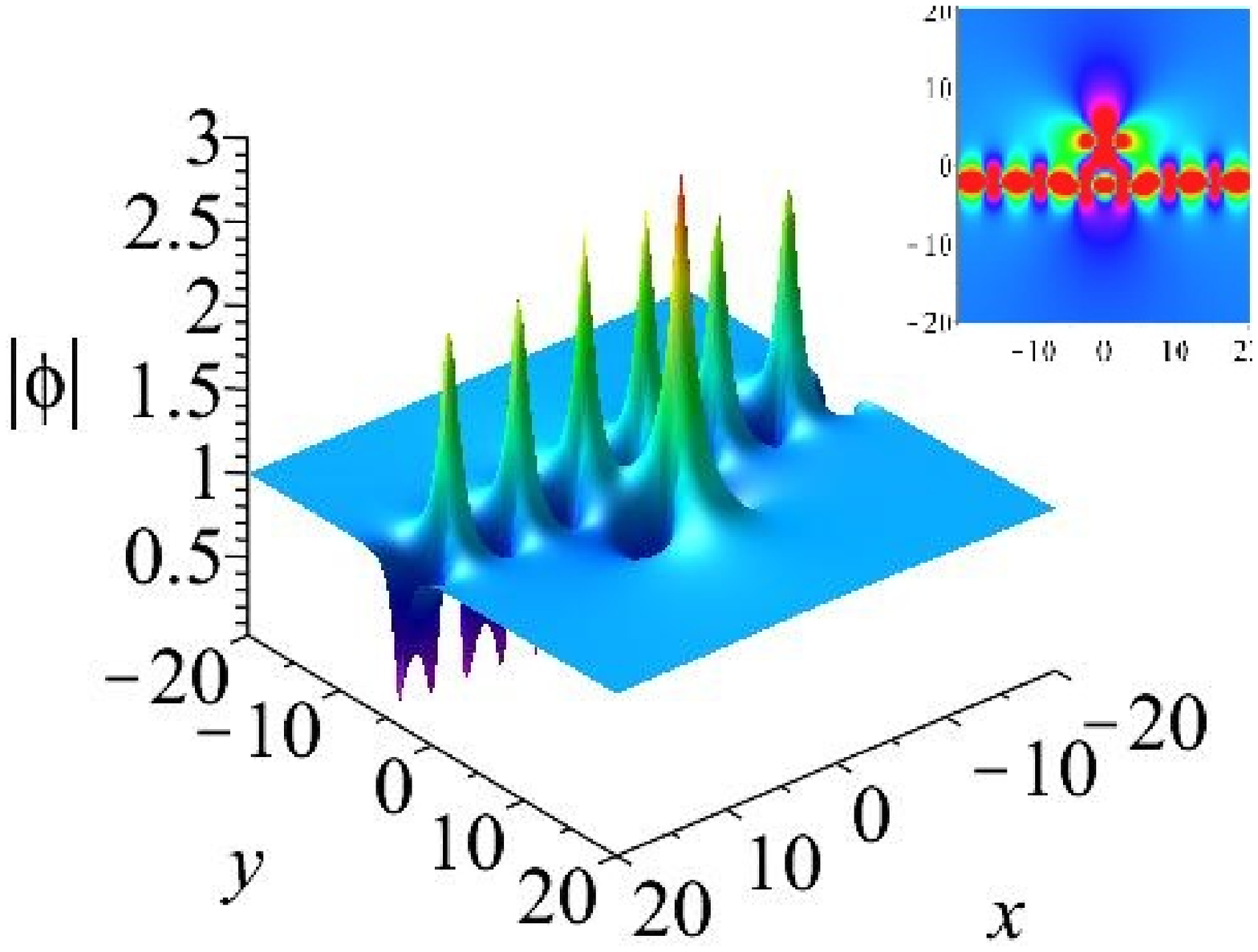}}
\subfigure[$\eta_{3}^0=\pi$]{\includegraphics[width=5cm]{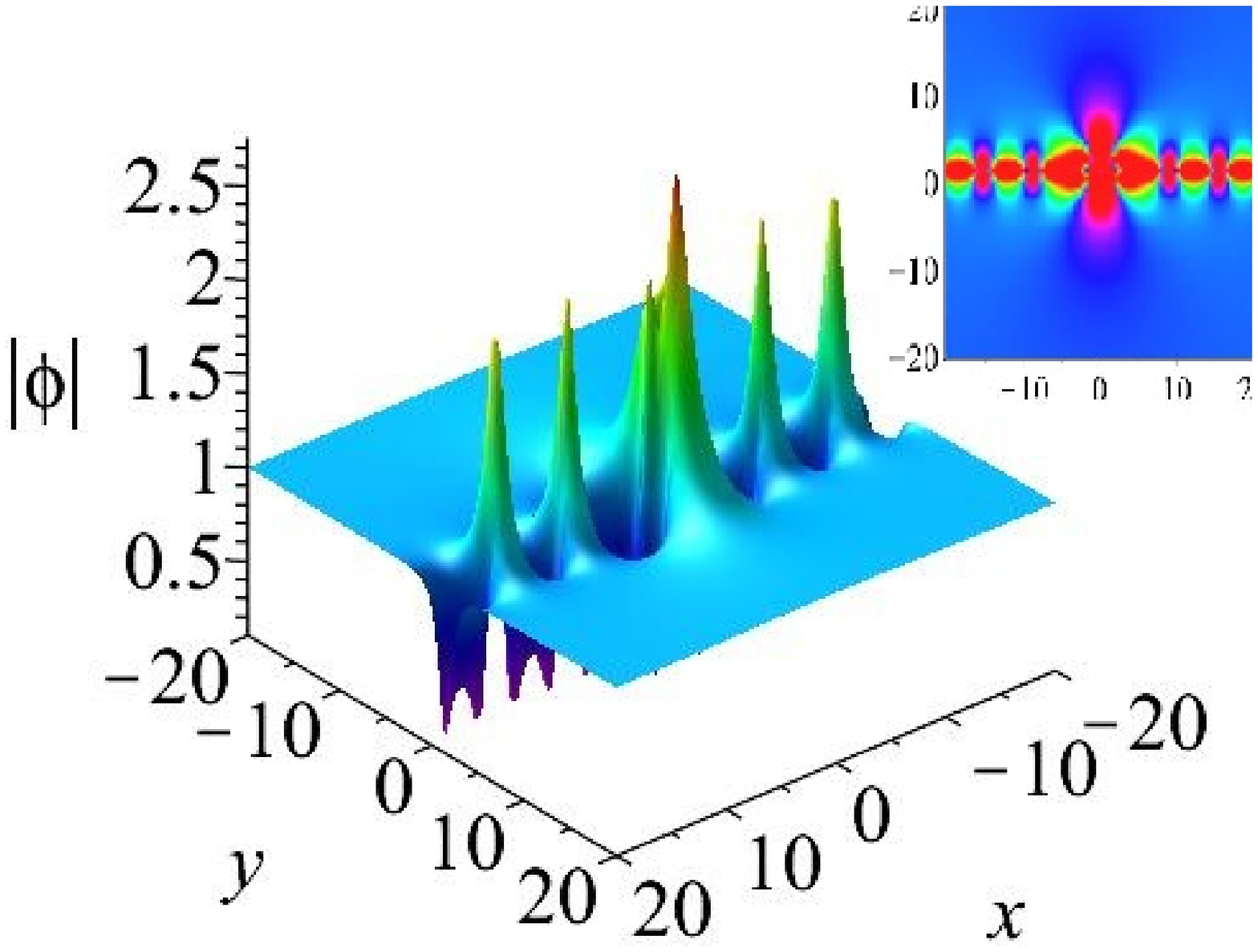}}
\subfigure[$\eta_{3}^0=2\pi$]{\includegraphics[width=5cm]{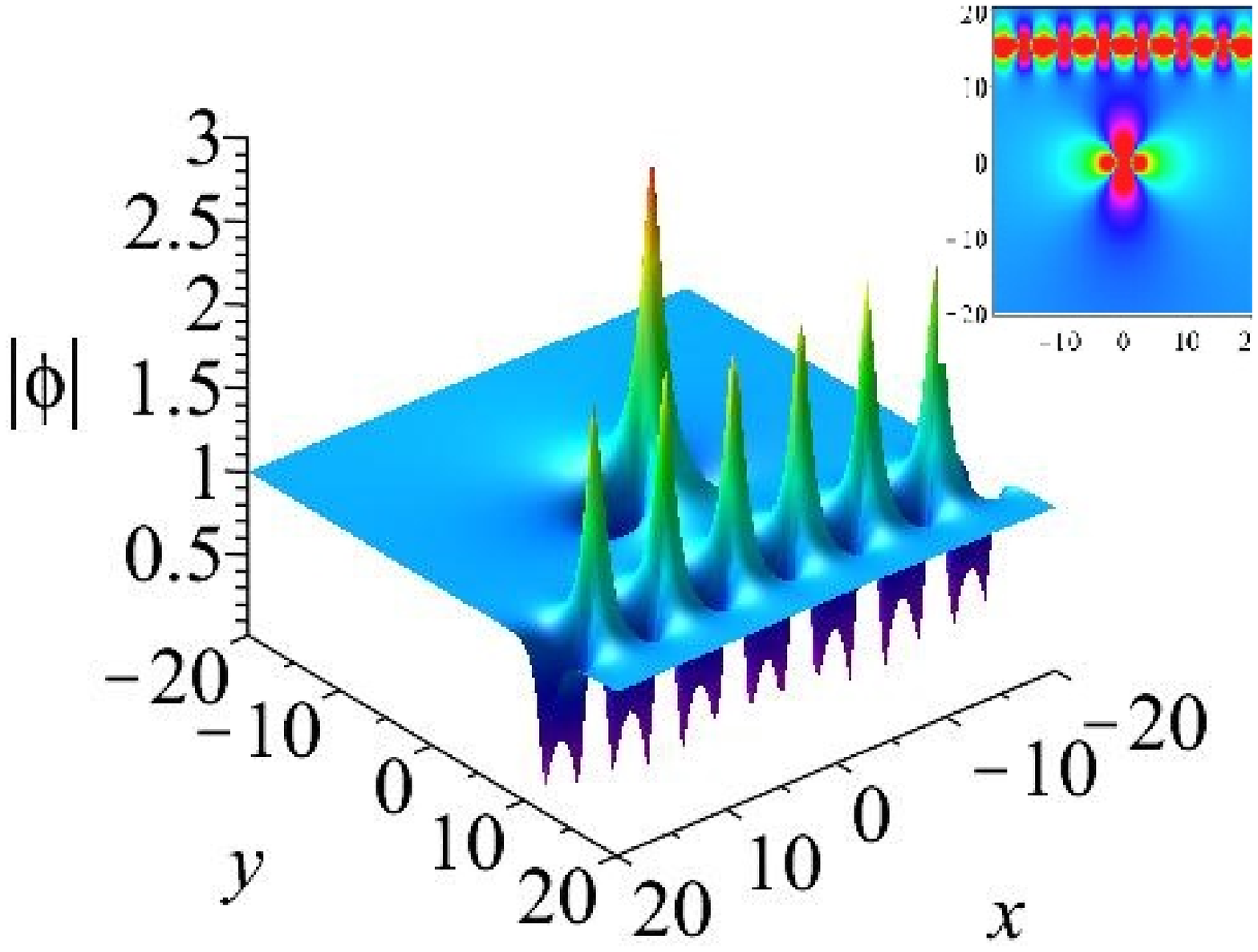}}
\caption{The superposition of a lump  and one-breather with parameters $\kappa=1,\lambda_1=1,\lambda2=-1,P_3=-P_4=1, Q_3=Q_4=1,t=0$ and different parameters $\eta_3^0$.~}\label{fig12}
\end{figure}

\noindent\textbf{Case 2: A hybrid of breathers and periodic line waves }$\\$
Another type of semi-rational solutions is the mixed solutions consisting of lumps and breathers. Here we only consider the simplest one of this type of mixed solution, which only possesses one lump and one breather.
To this end, we take parameters in \eqref{rfg}
\begin{equation} \label{pa-6}
\begin{aligned}
N=4\,,Q_{1}=\lambda_{1}P_{1}\,,Q_{2}=\lambda_{2}P_{2}\,,\exp(\eta_{1}^{0})=\exp(\eta_{2}^{0})=-1\,,
\end{aligned}
\end{equation}
and take a limit as $$P_{1}\,,P_{2}\rightarrow 0,$$
then functions $f$ and $g$ of solutions can be rewritten as
\begin{equation} \label{hy-11}
\begin{aligned}
f=&e^{A_{34}}(a_{13}a_{23}+a_{13}a_{24}+a_{13}\theta_{2}+a_{14}a_{23}+a_{14}a_{24}+a_{14}\theta_{2}+a_{23}\theta_{1}+\\
&a_{24}\theta_{1}+\theta_{1}\theta_{2}+a_{12})e^{\eta_{3}+\eta_{4}}+(a_{13}a_{23}+a_{13}\theta_{2}+a_{23}\theta_{1}+\theta_{1}\theta_{2}+a_{12})e^{\eta_{3}}\\
&+(a_{14}a_{24}+a_{14}\theta_{2}+a_{24}\theta_{1}+\theta_{1}\theta_{2}+a_{12})e^{\eta_{4}}+\theta_{1}\theta_{2}+a_{12}\,,\\
g=&e^{A_{34}}[a_{13}a_{23}+a_{13}a_{24}+a_{13}(\theta_{2}+b_{2})+a_{14}a_{23}+a_{14}a_{24}+a_{14}(\theta_{2}\\
&+b_{2})+a_{23}(\theta_{1}+b_{1})+a_{24}(\theta_{1}+b_{1})+(\theta_{1}+b_{1})(\theta_{2}+b_{2})+a_{12}]e^{\eta_{3}+i\phi_{3}+\eta_{4}+i\phi_{4}}\\
&+[a_{13}a_{23}+a_{13}(\theta_{2}+b_{2})+a_{23}(\theta_{1}+b_{1})+(\theta_{1}+b_{1})(\theta_{2}+b_{2})+a_{12}]e^{\eta_{3}+i\phi_{3}}\\
&+[a_{14}a_{24}+a_{14}(\theta_{2}+b_{2})+a_{24}(\theta_{1}+b_{1})+(\theta_{1}+b_{1})(\theta_{2}+b_{2})+a_{12}]e^{\eta_{4}+i\phi_{4}}\\
&+(\theta_{1}+b_{1})(\theta_{2}+b_{2})+a_{12},
\end{aligned}
\end{equation}
where $$a_{sl}= \frac{4P_l^3}{P_l^4+(\lambda_2P_l-Q_l)^2}\,(s=1,2,l=3,4),$$  and $a_{12}\,,b_{s}\,,\phi_{l},\eta_{l}\,,e^{A_{34}}$ are given by \eqref{rt} and \eqref{cs1}. Further, taking  parameters
\begin{equation} \label{co-hy-2}
\begin{aligned}
\lambda_{1}=-\lambda_{2}\,,P_{3}=-P_{4}\,,Q_{3}=Q_{4}\,,
\eta_{3}^{0}=\eta_{4}^{0*},
\end{aligned}
\end{equation}
thus semi-rational solutions consisting of a fundamental lump and one breather are obtained, see Fig.\ref{fig10}. The period of this breather is $\frac{2\pi}{P_3}$. In this situation, solution $u$ is real. Note that for fixed parameters $\eta_3^0,\eta_{4}^0$, the distance between the lump and the breather does not alter during their propagation in the $(x,y)$-plane. In order to observe the interaction between lump and breather,  we control the location of the breather by altering parameters $\eta_3^0,\,\eta_4^0$.  As can be seen in Fig.\ref{fig12}, when $|\eta_3^0|\rightarrow0$, the distance between the lump and the breather ten to zero. The lump immerses
in breather (see the panel of $\eta_3^0=0$).  The
superimposition of lump with breather excites a peak whose height is less than three times of the
background amplitude.
Besides, the breather possesses a lower amplitude and a smaller period than Fig. \ref{fig10} and it propagates
stably. Apparently, the wave structure of the lump is destroyed due to the interaction, and energy transfer has
happened between lump and breather.

Higher-order semi-rational solutions consisting of higher-order lumps and higher-order breathers can also be generated by a similar way with larger $N$, the related results will be published elsewhere.

\section{Summary and discussion}\label{con}
In this paper,  we have introduced and investigated an partial reverse space-time nonlocal Mel'nikov equation, which is an multidimensional versions of the nonlocal Schr\"odinger-Boussinesq equation with  a parity-time-symmetric potential. By using the Hirota's bilinear method, soliton solutions are obtained. Although these soliton solutions
have singularities, general $n$-breather solutions and mixed solutions consisting of breathers and periodic line waves can be derived under proper parameter constraints, which are nonsingular, see Figs.\ref{fig1}, \ref{fig11}, \ref{fig2}.  Taking a long wave limit of these obtained solitons and parameter constraints, nonsingular rational solutions have be generated, which are lump solutions. The exact explicit lump
solutions up to the third-order are presented, dynamical behaviours of interaction between lumps have been demonstrated, see Figs. \ref{fig3},\ref{fig4}, \ref{fig41}.
Besides, taking a lone wave limit of solitons partially,  two subclasses of semi-rational solutions are derived. One subclass of these semi-rational solutions describe lumps on a background of periodic line waves, see Fig. \ref{fig5}, \ref{fig6}, \ref{fig9}. Another one describes interaction between lumps and breathers, see Figs.\ref{fig10}, \ref{fig12}.
In particular, two solutions of the nonlocal Schr\"odinger-Boussinesq equation, namely, fundamental rogue waves and mixed solution consisting of a rogue wave and periodic line waves are obtained as reductions of the corresponding
solutions of the partial reverse space-time nonlocal Mel'nikov equation.
\section*{Acknowledgment}
This work is sponsored by the National Natural Science Foundation of China (No. 11571079), Shanghai Pujiang Program (No. 14PJD007)
and the Natural Science Foundation of Shanghai (No. 14ZR1403500 ), and the Young
Teachers Foundation (No. 1411018) of Fudan university.
The author Liu thanks Prof. Jingsong He of NingBo university and Dr. Jiguang Rao of USTC for their many discussions and suggestions on the paper.

%\begin{acknowledgements}
%\end{acknowledgements}



%%%%%%%%%%%%%%%%%%%%%%%%%%%%%%%%%%%%

%\begin{thebibliography}{}
% etc
%\end{thebibliography}

\end{document}